\documentclass{aa} 
\usepackage[utf8]{inputenc}
\usepackage{soul}
\usepackage{graphicx}
\usepackage{amsmath, amssymb}
\usepackage{subcaption}
\usepackage{makecell}
\usepackage{multirow, array}
\usepackage[pdfencoding=auto, psdextra]{hyperref}
\usepackage[normalem]{ulem}
\usepackage[switch]{lineno}
\usepackage{comment}

\newcommand{\Fermi}{{\sl Fermi}}

\hypersetup{
    colorlinks=true,
    linkcolor=blue,
    filecolor=magenta,      
    urlcolor=cyan,
    }

\newcommand{\g}{$\gamma$}

\begin{document}

\title{The First Deep Learning all-sky List of \Fermi-LAT Gamma-ray Sources}

\author{S. Bhattacharyya\inst{\ref{inst1}}\thanks{corresponding author} 
\and G. Principe\inst{\ref{inst3}}\inst{,\ref{inst4}}$^*$
\and D. Malyshev\inst{\ref{inst2}}$^*$ 
\and R. Ruiz de Austri\inst{\ref{inst5}}
\and G. Zaharijas\inst{\ref{inst1}}
\and S. Caron\inst{\ref{inst6}}\inst{,\ref{inst7}}
\and F. Stoppa\inst{\ref{inst7}}\inst{, \ref{inst8}}
\and G. Johannesson\inst{\ref{inst9}}
}

\institute{University of Nova Gorica, Centre for Astrophysics and Cosmology,
Vipavska 11c, Ajdovščina, Slovenia\label{inst1}
\and Istituto Nazionale di Fisica Nucleare, Sezione di Trieste, I-34127 Trieste, Italy \label{inst3}
\and Dipartimento di Fisica, Universitá di Trieste, I-34127 Trieste, Italy \label{inst4}
\and Erlangen Centre for Astroparticle Physics, Nikolaus-Fiebiger-Str. 2, 91058, Erlangen, Germany \label{inst2}
\and Instituto de Física Corpuscular, IFIC-UV/CSIC, Valencia, Spain\label{inst5}
\and Nikhef, Science Park 105, 1098 XG Amsterdam, the Netherlands\label{inst6}
\and Radboud University, Department of High Energy Physics,
Heyendaalseweg 135, Nijmegen, The Netherlands\label{inst7}
\and Department of Physics, University of Oxford, Denys Wilkinson Building, Keble Road, Oxford OX1 3RH, UK\label{inst8} 
\and CCP Games, Reykjavik, Iceland\label{inst9}
}

\date{Received date /
Accepted date}

\abstract{Context: 
State-of-the-art methods for the detection and characterization of \g-ray sources rely on likelihood-based analysis.
However, large and uncertain diffuse \g-ray backgrounds, particularly near the Galactic plane, present a major challenge to these approaches.

Aims: 
We develop a deep-learning-based pipeline to construct a list of $\gamma$-ray sources and to quantitatively assess the impact of background uncertainties on source detection and characterization. 
This work extends and enhances the existing AutoSource-ID algorithm, previously adopted on optical observations, for applications on \g-ray data.

Methods:
Source detection was performed through image segmentation using a U-Net architecture, while dedicated deep neural networks (DNNs) were employed for source localization, flux estimation, and the classification of detected sources as either genuine or spurious. The training data were generated from Monte Carlo simulations based on 10 years of {\Fermi}-LAT observations. To produce realistic representations of the \g-ray sky, we adopted the source properties from the fourth {\Fermi}-LAT source catalog data release 2 (4FGL-DR2), together with the corresponding diffuse emission model.

Results: 
We present the first {\Fermi}-LAT \g-ray all-sky source list derived using a deep-learning-based method (1FDL), containing more than 3800 sources with estimated positions, fluxes, and probability scores that quantify the likelihood of each detection being genuine or spurious. We further evaluated the robustness of the pipeline against variations in the interstellar emission model (IEM) and compared the resulting source list with the 4FGL-DR2 catalog.

Conclusion: 
We find that the our deep-learning-based approach achieves performance comparable to, or better than, the likelihood-based methods employed in the standard {\Fermi}-LAT catalog pipeline when tested against changes in the IEM. The 1FDL source list is broadly consistent with the {\Fermi}-LAT 4FGL-DR2 catalog obtained over the same 10-year observational period, including a few new sources confirmed in the recent {\Fermi}-LAT 16-year source list (FL16Y). These results demonstrate the potential of DNN methods for the detection and characterization of \g-ray sources and provide a foundation for their application in future high-energy astrophysical analyses.
}


\maketitle
\tableofcontents

\section{Introduction}

Recent advances in deep neural networks (DNNs) have transformed many areas of science and engineering, yet their application to the detection and characterization of $\gamma$-ray sources remains
scarce~\citep{2020AN....341..819M, Boris2021, 2025PhRvD.112j3022E}.
One reason for this is the success of
traditional  likelihood-based methods. In particular, parameter estimation through maximum likelihood estimation (MLE) is statistically well motivated: under the assumption that the adopted model correctly describes the data, MLE provides asymptotically efficient and unbiased estimators. According to the Cramér--Rao bound \citep{cramer:stat, Rao1992:stat}, no unbiased estimator can achieve smaller statistical uncertainties than those obtained with MLE.

In practice, however, the performance of MLE strongly depends on the adequacy of the assumed model. 
In $\gamma$-ray astronomy, one of the major challenges is the imperfect modeling of the diffuse Interstellar Emission Model \citep[IEM,][]{Strong:1998fr, 2007ARNPS..57..285S, Vladimirov:2010aq, Acero2016Template}.
The diffuse Galactic $\gamma$-ray background models are affected by substantial uncertainties related to the distribution of interstellar gas, particularly toward the Galactic center, as well as to the spatial distribution of cosmic-ray (CR) sources, CR propagation, and energy losses in Galactic radiation fields. As a consequence, mismodeling of the diffuse emission can produce residual structures that may be incorrectly interpreted as point-like sources, leading both to spurious detections and to biased reconstruction of source properties. This remains one of the major unresolved systematic uncertainties in high-energy astrophysics.

A practical approach to estimate this modeling uncertainties is to compare the results obtained with different IEMs. However, these models are generally based on the same underlying observational data, such as surveys of Galactic gas tracers, so systematic inaccuracies in the input data propagate to all derived IEMs. More fundamentally, the traditional MLE framework is limited by the flexibility of the adopted sky model, which may not be sufficiently expressive to capture the full complexity of the diffuse emission. 

As universal approximators, DNNs have the advantage that can learn complex, non-linear representations directly from the data and may therefore be less sensitive to mismodeling of the diffuse background. In general, point-source detection methods can be divided into two classes: methods that explicitly model the diffuse emission \citep[e.g., in the DNN context,][]{2025PhRvD.112j3022E}, and methods that attempt to suppress or filter it out \citep[e.g.,][]{Boris2021}.
One of the advantages of filtering models is that they can be used to distinguish point-like sources from extended emission based on some general features, such as the correlation in the extended emission on angular scales larger than the size of the PSF.
Such methods may therefore be intrinsically more robust against uncertainties in the foreground emission. A classical example is the wavelet transform, which was successfully employed in the construction of the first {\Fermi}-LAT catalog of low-energy sources, 1FLE~\citep{fle-giacomo}.

In this paper, we use deep learning to directly detect point-like sources without explicitly modeling the diffuse foreground emission. Unlike traditional wavelet-based methods, where the filtering kernels are predefined, the filtering transformation in our method is learned automatically during the training process.

In our previous work \citep{Boris2021}, we introduced the AutoSourceID (ASID) pipeline, a machine learning (ML) tool for source detection and classification on simulated gamma-ray maps. In that proof-of-principle study, we demonstrated that the source detection threshold for simulated high-latitude sources ($|b|>20^{\circ}$) was comparable to that of the 4FGL catalog. However, the analysis was performed on simplified simulated data containing only two source classes, the energy dependence of the PSF was neglected, and the pipeline was not applied to real {\Fermi}-LAT observations. 

In this work, we i) build a more optimal pipeline that takes into account the energy dependence of the point spread function (PSF);
ii) improve the realism of training data by using four different source classes;
iii) extend the characterization stage to include dedicated networks for flux estimation and source validation;
iv) investigate the robustness of the pipeline against uncertainties in the diffuse background by testing its performance under different IEMs; in order to test the effect of uncertainties in the IEM, we train the DNN with data generated using one model and test it with the data either generated with the same or with a different model, 
we compare the performance of our pipeline with 4FGL-DR2 catalog~\citep{4fgl2020, Ballet:2020hze};
v) apply the pipeline to 10 years of {\Fermi}-LAT observations and construct the first deep-learning-based list of $\gamma$-ray sources, denoted 1FDL, using the same observational period as the 4FGL-DR2.
To date, this is the first study to directly compare the performance of a DNN-based pipeline with that of a likelihood-based pipeline for gamma-ray source detection and localization. The whole pipeline, including codes and environment management, and the 1FDL list are publicly available on GitHub\footnote{\href{https://github.com/suvoooo/Fermi-LATV2-DL}{GitHub Link} \includegraphics[height=4.0mm]{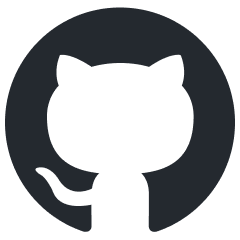}}.

The paper is organized as follows. In Section \ref{sec:train-data}, the data selection and the Monte Carlo generation of training data are described. Section \ref{sec:detection} discusses the source detection and localization, including the performance estimates on the simulated data. We present the source characterization module, including flux prediction, classification and location refinement processes and results on the simulated data in Section \ref{sec:CharModule}. We discuss the robustness of our pipeline against changing IEM in Section \ref{sec:bkg-syst}. In Section \ref{sec:realdata}, we apply the algorithm to the real data and construct the first catalog of gamma-ray sources based on deep learning.
In Section \ref{sec:conclusions}, conclusions and future prospects are discussed.
Additional technical details regarding our pipeline implementation are provided in the appendices.


\section{Data Selection and Preparation}
\label{sec:train-data}

The \textit{Fermi} Large Area Telescope (LAT) \citep{Atwood2009} has been continuously surveying the high-energy $\gamma$-ray sky since August 2008, detecting photons in the energy range from $\sim 20$ MeV to beyond 300 GeV. The fourth {\Fermi}-LAT source catalog \citep[4FGL,][]{4fgl2020} and its second data release \citep[4FGL-DR2,][]{Ballet:2020hze}, based on 8 and 10 years of observations respectively, contain 5064 and 5700 detected sources.
The majority of the detected sources are of extragalactic origin and are predominantly associated with active galactic nuclei (AGNs), in particular BL Lacertae objects (BLLs) and flat-spectrum radio quasars (FSRQs). Among Galactic sources, pulsars (PSRs) constitute the largest population, with additional significant contributions from pulsar wind nebulae (PWNe) and supernova remnants (SNRs).

For the generation of the training and test datasets, we used the \textrm{Fermitools}\footnote{https://fermi.gsfc.nasa.gov/ssc/data/analysis/software/}  (v2.0.8) software package together with the 4FGL-DR2 catalog (file version \texttt{gll\_psc\_v27.fit}\footnote{https://fermi.gsfc.nasa.gov/ssc/data/access/lat/10yr\char`_catalog/}). The simulations incorporate both Galactic and extragalactic source populations, as well as different diffuse IEMs, described in the following sections.

\subsection{Data Selection}
\label{sec2d1-Fermitools}

We use 522 weeks ($\sim 10$ years) of {\Fermi}-LAT observations collected between August 4, 2008 and August 5, 2018, corresponding to the same time interval adopted for the construction of the 4FGL-DR2 catalog. Photons were selected in the energy range between 300 MeV and 1 TeV. The lower energy threshold was chosen to reduce source confusion and contamination from the diffuse background at low energies \citep{fle-giacomo}.

In this work, we consider only \textrm{`FRONT'} (\textrm{`F'} from now) converting events, which have a better angular resolution at a given energy compared to \textrm{`BACK'} (\textrm{`B'} from now) converting events. In order to take advantage of the better angular resolution of LAT at higher energies \citep{Ajello-2021-LAT-Performance}, we adopt different spatial resolutions for the six energy bins used in this analysis (Table~\ref{tab:IRFs}). Using six energy bins is a balance between resolving the detector performance and overly complicating the analysis pipeline.
We apply zenith angle cuts of $105^{\circ}$ for photons with energies above 1 GeV and $100^{\circ}$ for photons below 1 GeV in order to suppress contamination from the Earth's limb. Source-class events processed with the \textrm{P8R3\_SOURCE\_V2} \citep{bruel2018} instrument response functions (IRFs) are used throughout this work, corresponding to the event selection available at the time the training data were generated.

We use the standard quality selection criteria \textrm{(DATA\_QUAL $== 1$) \&\& (LAT\_CONFIG $== 1$)} and calculate the livetime cube using \textrm{`gtltcube'} in \textrm{Fermitools}. 
We derive the binned exposure maps using \textrm{`gtexpcube2'}, while the maps for point sources and IEMs are generated with \textrm{`gtsrcmaps'} and \textrm{`gtmodel'}, respectively. We include the energy dispersion via the keyword \textrm{edisp\_bins=-2} in the steps above. 
Finally, Poisson fluctuations are applied to the simulated count maps in order to reproduce the statistical properties of the observed data.

\begin{table}
\caption{The spatial resolution for creating sky-maps in the energy bins used in our analysis. 
For the sky maps, we use the healpy version~\citep{healpy} of the Healpix package~\citep{2005ApJ...622..759G}. H, W, C refer to the height, width and number of channels in an image.}
\label{tab:IRFs}
\centering
\begin{tabular}{cccc}
\hline\hline
Energy Bin & Healpix & Resolution & Patch Size\\
 (GeV) &  $\log _2(\text{Nside})$ & deg. & (H=W; C=1)\\
\hline
0.3-0.5 & 7  & $\approx 0.4^{\circ}\, $ & $32\, $\\
0.5-1.0 & 7  & $\approx 0.4^{\circ}\, $ & $32\, $ \\
1.0-2.0 & 8  & $\approx 0.2^{\circ}\, $ & $64\, $ \\
2.0-7.0 & 9  & $\approx 0.1^{\circ}\, $ & $128\, $ \\
7.0-20.0 & 10  & $\approx 0.06^{\circ}\, $ & $256\, $\\
20.0-1000 & 10  & $\approx 0.06^{\circ}\, $ & $256\, $\\
\hline
\end{tabular}
\end{table}

For the simulation of the sky maps, we use the source properties of the main source classes in the 4FGL-DR2 catalog: BLLs, FSRQs, Blazars of Uncertain types (BCUs), pulsars (PSRs) and the class PWN-SNR, which includes pulsar wind nebulae (PWNe), supernova remnants (SNRs) and sources of unknown nature that overlap with known SNRs or PWNe (SPP). We note that millisecond pulsars are included in PSRs in 4FGL-DR2 catalog but are a separate class starting with the 4FGL-DR3 catalog \citep{4FGL-DR3}.

\subsection{Distribution of Spectral Parameters}

We describe BLLs, FSRQs, and PWN-SNR sources with a log-parabola (LP) spectrum,
\begin{equation}\label{eq:logparabola}
\frac{dN}{dE} = K\left(\frac{E}{E_0}\right)^{-\alpha - \beta, \log(E/E_0)},
\end{equation}
where $K$ is the normalization at the pivot energy $E_0$, and $\alpha$ and $\beta$ are the spectral slope and curvature, respectively. BCUs are assigned to either the BLL or FSRQ class according to their relative abundances in 4FGL-DR2 and are modeled with the same LP form. The parameters $\alpha$, $\beta$, and $\log_{10}(K)$ are sampled from Gaussian distributions derived from the 4FGL-DR2 catalog, while $E_0$ is modeled with a log-normal distribution. Although a strong correlation between $\alpha$ and $E_0$ is observed in the \Fermi~catalog, this correlation was not taken into account during the sampling process. An example of LP spectral parameter distribution is shown in Figure \ref{fig:bll-params}.

For PSRs, we adopt a power law with exponential cutoff (PLEC),
\begin{equation}\label{eq:plec}
\frac{dN}{dE} = K\left(\frac{E}{E_0}\right)^{-\Gamma}
\exp\left[a\left(E_0^b - E^b\right)\right],
\end{equation}
where $\Gamma$ is the low-energy spectral index, $b$ the cutoff index, and $a$ the exponential factor in units of $\mathrm{MeV}^{-b}$. The parameters $\Gamma$, $b$, and $\log_{10}(K)$ are modeled with Gaussian distributions with the mean and variance derived as the sample mean and sample variance, while $E_0$ and $a$ follow log-normal distributions.
Following the 4FGL convention, we impose $\Gamma \geq 0$ \citep{4fgl2020}. An example of PLEC parameter distribution is shown in Figure \ref{fig:plec-params}.

\subsection{Spatial Distribution}

The spatial distribution of extragalactic sources (blazars) is assumed to be isotropic. Source positions are generated by sampling the Galactic longitude uniformly in the range $[0,2\pi]$ and $\sin b$ uniformly in the interval $[-1,1]$.
The spatial distribution of PSRs is modeled with Gaussian components fitted to the 4FGL-DR2 catalog. The longitude distribution is described by a single Gaussian, while the latitude distribution is modeled as a double Gaussian centered on the Galactic plane in order to account for the two main pulsar populations: young pulsars, concentrated near the plane, and millisecond pulsars, which exhibit a broader latitude distribution due to their migration away from the plane.
For the PWN-SNR class, both the latitude and longitude distributions are modeled with single Gaussian functions fitted to the corresponding 4FGL-DR2 source population.

\subsection{Energy Flux Distribution}

In the mock catalogs, the energy flux distribution of each simulated source class is constructed to reproduce the corresponding \textrm{Energy\_Flux100} distribution in the 4FGL-DR2 catalog down to the flux bin containing the maximum number of sources. Below this peak, we adopt a flat extrapolation in logarithmic flux space, following \cite{Boris2021}, since our aim is to explore the possibility of detecting faint sources below the LAT detection threshold. This extrapolated component, denoted \textrm{F0}, extends one additional flux bin beyond the lowest populated catalog bin.

\begin{figure}
    \centering
    \includegraphics[width=0.9\linewidth]{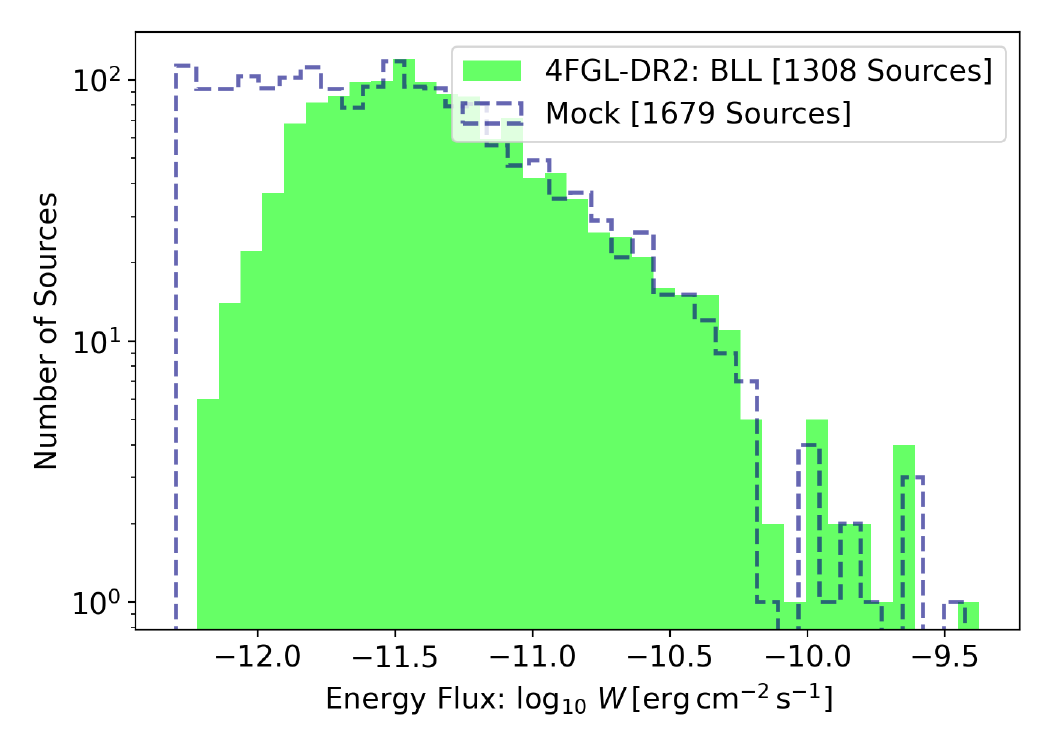}
    \caption{$\log 10$ of the Energy Flux distribution used for data generation (blue) is shown here for `BLL' source only compared to the distribution in the DR2 catalog (green).
    }
    \label{fig:Int-Flux-Train-Data}
\end{figure}

Candidate mock sources are generated by sampling spectral and spatial parameters from the distributions described above. Their corresponding \textrm{Energy\_Flux100} values are then computed and assigned to flux bins. For bins above the peak of the observed distribution, sources are accepted provided that the cumulative number of simulated sources does not exceed the corresponding 4FGL-DR2 count by more than $10\%$. For bins below the peak, the same acceptance criterion is applied using the extrapolated source counts. An example of the resulting energy flux distribution for simulated BLL sources is shown in Figure~\ref{fig:Int-Flux-Train-Data}.  

\subsection{Diffuse Emission Model}

The main challenge in detecting point sources, particularly faint ones, lies in distinguishing them from the diffuse IEM. This difficulty is most severe below 1 GeV and near the Galactic plane, especially within the Galactic ridge ($|l| < 60^{\circ}$), where the diffuse emission is both bright and highly structured due to the distribution of interstellar gas \citep{4fgl2020}. Uncertainties in the IEM arise from imperfect knowledge of the gas distribution, the interstellar radiation field (ISRF), and spatial variations in the cosmic-ray density.

As our benchmark diffuse model, we use the latest {\Fermi}-LAT IEM, \texttt{gll\_iem\_v07.fits}\footnote{https://fermi.gsfc.nasa.gov/ssc/data/access/lat/}, together with the isotropic template \texttt{iso\_P8R3\_SOURCE\_V3\_v1.txt}, which accounts for unresolved extragalactic emission and the residual cosmic-ray background. We refer to this configuration as B1.
To assess the robustness of the pipeline against diffuse-model uncertainties, we also use an alternative IEM, \texttt{gll\_iem\_v06.fits}, originally adopted in the construction of the 3FGL catalog \citep{3fglAcero} and rescaled for Pass 8 Source class events. This alternative model is denoted B2. Both models were developed for {\Fermi}-LAT catalog analyses and tuned to reproduce the observed data, although with different fitting procedures. Similar comparisons between diffuse models were performed in the development of the 4FGL catalog \citep{4fgl2020}.


\section{Source Detection and Localization}
\label{sec:detection}

\begin{figure*}
    \centering
    \includegraphics[width=0.94\textwidth]{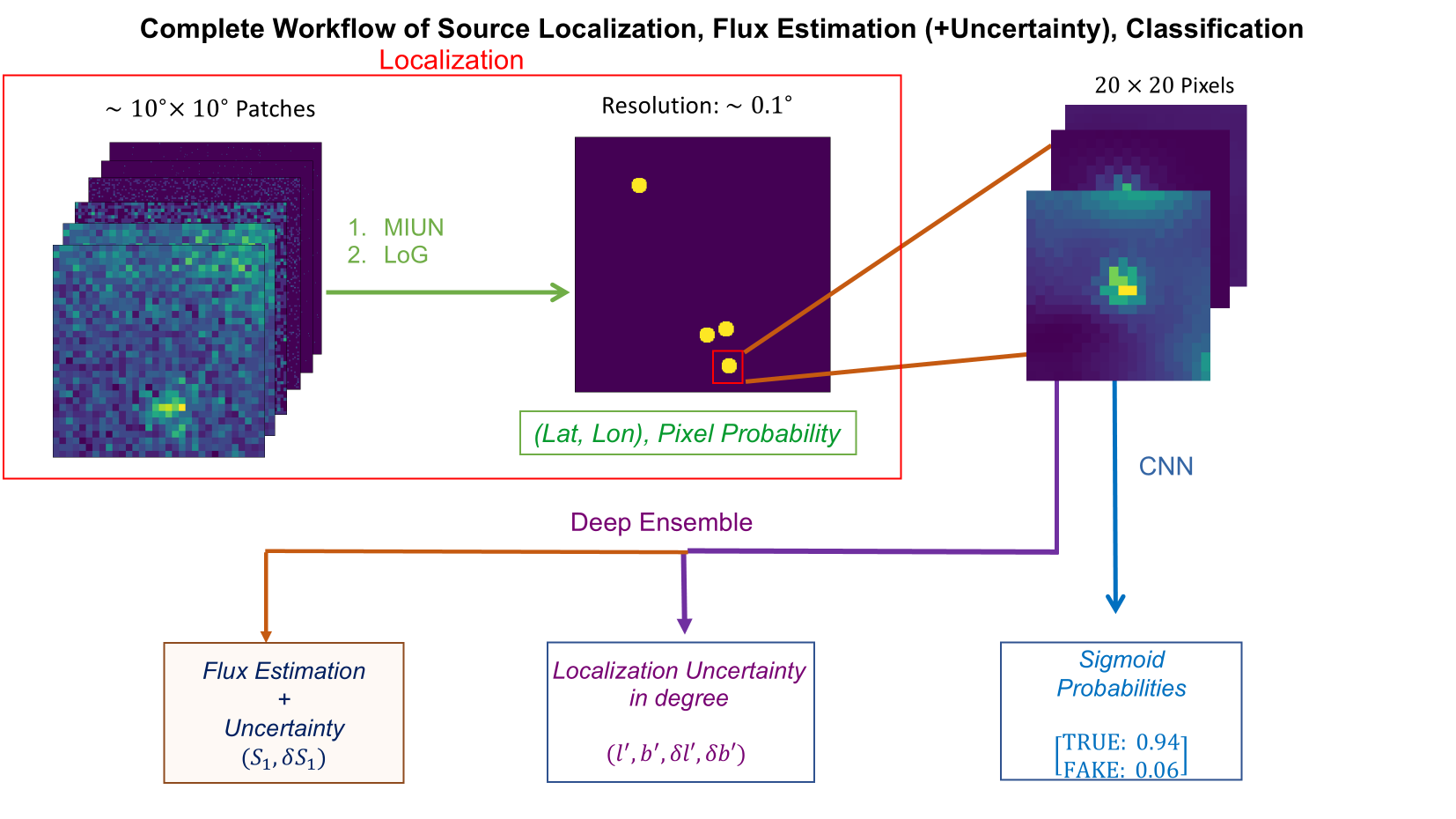}
    \caption{The complete workflow for catalog generation in ASID-1 is shown here. The first step involves source detection and localization via segmentation (with a Multi-Input UNET i.e. MIUN) and source center estimation with LoG. Smaller patches $(20\times 20)$ are then generated around the predicted source location as the centre of the patch. They are used for three separate tasks: per-source classification, per-source refined localization with uncertainty prediction and per-source flux estimation. More details are in the text.}
    \label{fig:workflow}
\end{figure*}

Our pipeline consists of two main modules (Figure~\ref{fig:workflow}):
\begin{enumerate}
    \item \textit{Detection \& Localization} module (described in this section), which forms the core of the pipeline and produces a list of source candidates together with their sky coordinates.
    
    \item \textit{Characterization} module (Section~\ref{sec:CharModule}), which further analyzes the detected candidates through three dedicated submodules:
    \begin{enumerate}
        \item \textit{Flux Estimation}: predicts the flux above 1~GeV and its uncertainty;
        \item \textit{True-Spurious Classification}: assigns a probability for a candidate to correspond to a genuine or spurious source;
        \item \textit{Location Refinement}: improves the source localization and estimates the associated uncertainty.
    \end{enumerate}
\end{enumerate}

The \textit{Detection \& Localization} module operates on $\sim 10^{\circ}\times10^{\circ}$ sky patches, whereas the \textit{Characterization} module uses smaller $\sim 2^{\circ}\times2^{\circ}$ regions centered on the candidate source positions.

\subsection{Source Detection with Semantic Segmentation}

A major challenge in astrophysical source detection is that the number of sources in an image is not known a priori, making standard image-classification approaches unsuitable. Instead, we employ semantic segmentation, where each pixel is classified as belonging either to a source or to the diffuse background. For this task, we use a network based on the U-Net architecture \citep{UNET}, which predicts source and background probabilities for every pixel in the input image.

The input image sizes differ across energy bins in order to reflect the energy dependence of the LAT angular resolution. Although lower-energy bins contain higher photon statistics, the positional information is substantially more accurate at higher energies with respect to lower energies. We therefore adopt the spatial resolution of the $2$--$7$ GeV bin, corresponding to $\sim 0.1^{\circ}$, for the segmented output images and for the localization procedure described in Section~\ref{sec:localization}.

The original U-Net consists of an encoder--decoder convolutional neural network (CNN) architecture. In the encoding path, the spatial dimensions are progressively reduced while the number of feature channels increases, followed by a symmetric decoding path that reconstructs the target segmentation map. Skip connections between encoder and decoder layers preserve small-scale spatial information and improve localization accuracy.

To simultaneously process images from multiple energy bins with different spatial dimensions, we extend the standard U-Net architecture into a Multi-Input U-Net (MIUN). In this framework, features are first extracted independently from images of different sizes and are subsequently combined within a common U-Net architecture (details are provided in Appendix~\ref{app:det-and-loc}). Although the network operates on multiple inputs, it produces a single segmented output image.

The training masks are generated by placing circular regions around the simulated source positions, with pixel values set to unity inside the circle and zero elsewhere. The disk radius $R$ is treated as a hyperparameter and optimized through dedicated tests; throughout this work we adopt $R=8$ pixels, corresponding to $\sim 0.5^{\circ}$ in the highest-energy bin ($7$--$20$ GeV).

The MIUN output consists of pixel-wise scores between 0 and 1. We train the network using the binary cross-entropy (BCE) loss and evaluate its performance with the Dice coefficient \citep{segmentation-dice},
\[
\mathrm{Dice} = \frac{2|X \cap Y|}{|X| + |Y|},
\]
where $X$ and $Y$ denote the true and predicted segmentation masks, respectively. To convert the predicted scores into binary masks, we scan thresholds between 0.1 and 1.0 in steps of 0.1 and find that a threshold of 0.4 provides the best performance. Pixels with scores above this threshold are classified as source pixels.

The network is trained using the Adam optimizer \citep{AdamOpt} for 150 epochs with a batch size of 128. The validation loss is monitored to adapt the learning rate and to implement early stopping, thereby reducing overfitting. The full pipeline is implemented in \textrm{TensorFlow}\footnote{\label{note1} https://www.tensorflow.org/}.

\subsection{Finding Source Center with Blob-Detection}
\label{sec:localization}

The output of the MIUN consists of segmentation maps of shape $(I,I,1)$, with $I=128$, containing pixel-wise source scores for each sky patch.

As discussed above, the segmented maps are constructed using the spatial resolution of the $2$--$7$ GeV energy bin, where the LAT PSF provides improved localization accuracy. Nevertheless, many sources do not produce sufficient photon statistics above $2$ GeV. The MIUN therefore combines information from lower-energy images, despite their broader PSF, to improve source detection performance. Once the segmented maps are obtained, source centers are identified in a second localization step.

For this purpose, we employ the Laplacian of Gaussian (LoG) algorithm, a standard blob-detection technique widely used in computer vision and astronomical source detection \citep{Fio-AsidL}. The LoG filter, also known as the Mexican-hat wavelet, was previously adopted in the PGWave algorithm~\citep{1997ApJ...483..350D} and used in several {\Fermi}-LAT analyses, including the 1FGL~\citep{2010ApJS..188..405A}, 2FGL~\citep{2012ApJS..199...31N}, and 1FLE~\citep{fle-giacomo} catalogs.

The LoG operator combines Gaussian smoothing with a Laplacian filter,
\begin{equation}
\label{eq:LoG}
\mathrm{LoG}(x,y;\sigma)=
\frac{1}{\pi\sigma^4}
\left[
\frac{x^2+y^2}{2\sigma^2}-1
\right]
\exp\left(-\frac{x^2+y^2}{2\sigma^2}\right),
\end{equation}
where the scale parameter $\sigma$ determines the characteristic blob size. Since the training masks are generated with a fixed source radius, they naturally provide an estimate of the optimal $\sigma$. The optimization procedure is described in Appendix~\ref{app:det-and-loc}.

Compared to our previous implementation based on the K-means clustering algorithm \citep{K-Means, Boris2021}, the LoG method is computationally more efficient, as it operates through convolution-like filtering rather than pixel-by-pixel clustering. This substantially accelerates the processing of the $128\times128$ segmented maps.

\subsection{Performance Evaluation}\label{sec:perform-eval-detect}

To quantify the performance of the detection pipeline, we use the standard precision (purity) and recall (completeness) metrics:
\begin{equation}
\text{Precision} = \frac{\text{TP}}{\text{TP} + \text{FP}}, \qquad \text{Recall} = \frac{\text{TP}}{\text{TP} + \text{FN}},
\label{eq:PreRec}
\end{equation}
where TP (true positive) denotes sources detected by the algorithm and present in the mock catalog, FP (false positive) denotes detected sources without a counterpart in the mock catalog, and FN (false negative) denotes mock-catalog sources missed by the algorithm. All quantities are evaluated as functions of source flux.

Following our previous work \citep{Boris2021}, a detected source is considered successfully associated with a mock-catalog source if their angular separation is smaller than $0.25^{\circ}$. 

We also note that the pixel resolution ranges from $\sim0.4^{\circ}$ in the lowest-energy bin (0.3--0.5 GeV) to $\sim0.06^{\circ}$ above 7 GeV. Consequently, for soft-spectrum sources with few photons above 2 GeV, the localization is driven primarily by the lower-energy data.
When multiple true sources lie within $0.25^{\circ}$ of a predicted source, the association is assigned to the source with the highest photon flux. Conversely, if multiple predicted sources are located within $0.25^{\circ}$ of a true source, the closest prediction is counted as a TP, while the remaining detections are kept in the pipeline for possible associations with fainter true sources. A detailed study of source confusion in close-pair configurations is presented in Appendix~\ref{app:det-and-loc}.

\begin{figure}
    \centering

    \includegraphics[width=0.94\columnwidth]{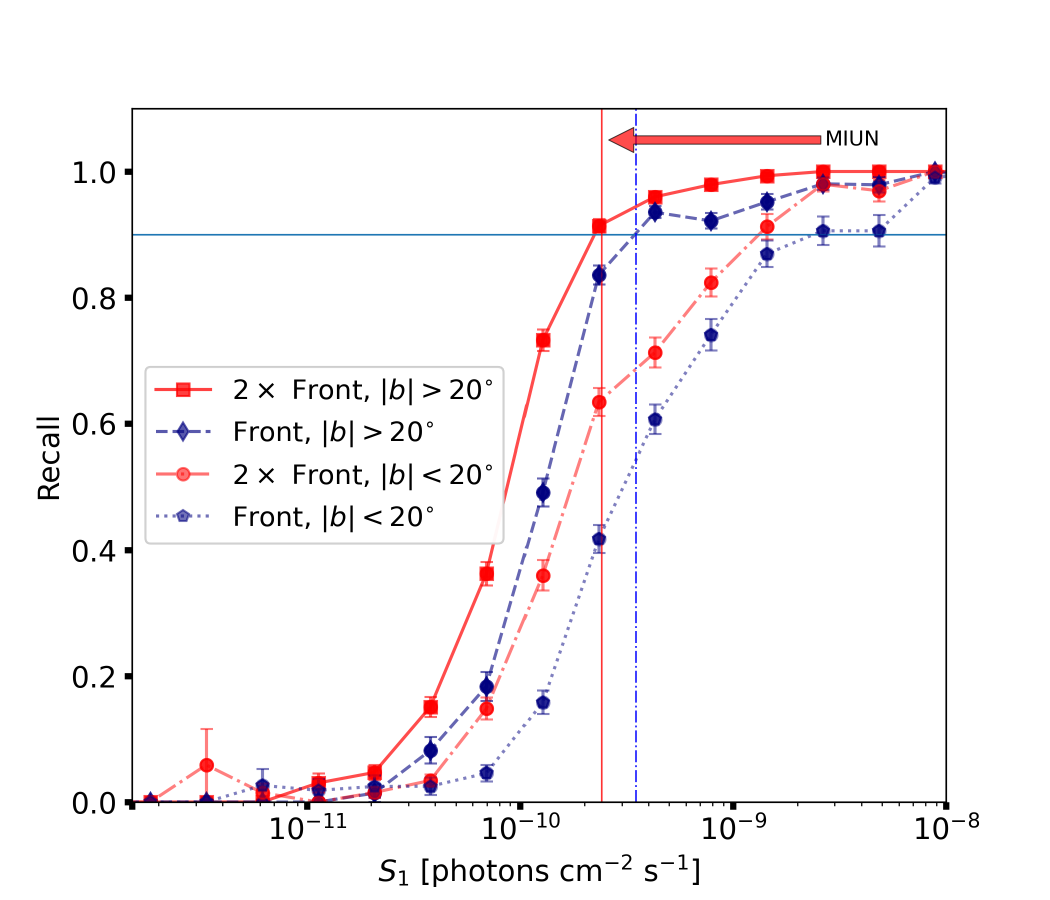}
    \caption{Recall (completeness) as a function of photon flux integrated above 1~GeV for high- and low-latitude sources, using \textrm{F} events only (blue) and doubled \textrm{F} statistics (red). The red vertical line marks the 90\% recall threshold of the MIUN pipeline for the doubled-\textrm{F} dataset. The blue dotted vertical line marks the same, but with \textrm{F} only dataset for $|b|>20^{\circ}$. Since the 4FGL sensitivity is based on combined \textrm{F+B} events, the F and doubled-\textrm{F} cases approximately bracket the expected \textrm{F+B} performance. For a comparison, the 4FGL photon flux sensitivity above 1 GeV is of $S_1 \sim 1.6 \times 10^{-10}\,\text{ph}\,\text{cm}^{-2}\,\text{s}^{-1}$. 
    }
    \label{fig:pre_rec_F0F1}
\end{figure}

Figure~\ref{fig:pre_rec_F0F1} shows the recall as a function of the photon flux integrated above 1~GeV ($S_1$) for high- ($|b|>20^{\circ}$) and low-latitude ($|b|<20^{\circ}$) sources. The uncertainty on the recall is estimated assuming a binomial distribution for TP and FN with fixed $N=\mathrm{TP}+\mathrm{FN}$,
\[
\sigma(\mathrm{Recall}) = \sqrt{\frac{p(1-p)}{N}},
\]
where $p=\mathrm{TP}/N$.
In the recall calculation, the simulated (true) values of $S_1$ are used for both TP and FN sources. This is not possible for FP sources, since no ground-truth flux exists for them. In Section~\ref{sec:flux_pred}, we introduce a dedicated flux-estimation network that predicts $S_1$ values for all detected candidates, enabling the evaluation of precision as a function of estimated flux.

We estimated the sensitivity of our method using a conservative criterion, defining it as the flux at which the recall derived from simulations reaches $90\%$ (see Fig.~\ref{fig:pre_rec_F0F1}). Assuming twice the \textrm{F} statistics, we obtain a sensitivity threshold of $\sim 2.3 \times 10^{-10}\,\text{ph}\,\text{cm}^{-2}\,\text{s}^{-1}$ ~$(\sim 2.4 \times 10^{-10}\,\text{ph}\,\text{cm}^{-2})$ for $|b|>20^{\circ}~ (|b|>10^{\circ})$. For comparison, the 4FGL catalog reports a sensitivity of $2 \times 10^{-12}\,\text{erg}\,\text{cm}^{-2}\,\text{s}^{-1}$ above 100~MeV for sources outside the Galactic plane ($|b| > 10^\circ$; \citealt{4fgl2020}). Assuming a power-law spectrum with photon index $\Gamma = 2.3$, close to the median spectral index in 4FGL, this corresponds to a photon flux of $S_1 \sim 1.6 \times 10^{-10}\,\text{ph}\,\text{cm}^{-2}\,\text{s}^{-1}$ above 1~GeV.
It should be noted, however, that the two sensitivity estimates are not defined in the same way. Our sensitivity is based on source recovery completeness, whereas the 4FGL sensitivity is inferred from the flux corresponding to the peak of the source-count distribution of detected sources (see Fig.~15 of the 4FGL catalog), which approximately marks the onset of catalog incompleteness. Therefore, the comparison should be interpreted with caution.
Moreover, our analysis is based exclusively on \textrm{F}-type events. Consequently, the threshold derived under the assumption of doubled \textrm{F} statistics should be regarded as a lower limit for a combined \textrm{F} + \textrm{B} analysis. For \textrm{F}-only data, the extragalactic source threshold is $\sim 3.6 \times 10^{-10}\,\text{ph}\,\text{cm}^{-2}\,\text{s}^{-1}$ ~$(\sim 4.0 \times 10^{-10}\,\text{ph}\,\text{cm}^{-2}\,\text{s}^{-1})$ for $|b|>20^{\circ}$ ~$(|b|>10^{\circ})$. Since back-converted LAT events have poorer angular resolution than front-converted events, the performance of a combined \textrm{F} + \textrm{B} analysis is expected to lie between the \textrm{F}-only and doubled-\textrm{F} cases shown in Fig.~\ref{fig:pre_rec_F0F1}.


\section{Source Characterization}
\label{sec:CharModule}

After source detection and localization, we proceed to the characterization stage. For each detected source, we use the LoG-derived pixel coordinates $(x,y)$ (Section~\ref{sec:localization}) to extract $20\times20$ pixel cutouts in each energy bin centered on the predicted source position. This patch size remains the same for all the modules in the source characterization pipeline. 

Because neighboring patches slightly overlap, the same source can appear in multiple cutouts. In such cases, we retain the cutout in which the source is closest to the patch center. For sources located near the edge of the larger $10^{\circ}\times10^{\circ}$ sky patch, the extracted regions are resized using bilinear extrapolation of the nearest edge values\footnote{https://scikit-image.org/docs/stable\\/auto\_examples/transform/plot\_rescale.html} to preserve the target $20\times20$ pixel shape ($\sim4~\mathrm{deg}^2$). This affects only $\sim0.7\%$ of patches $\geq 2$ GeV, increasing to $\sim13\%$ at energies below 1 GeV.

Moreover, because the recovery efficiency of closely spaced source pairs drops significantly for angular separations below $0.1^{\circ}$ (Appendix~\ref{app:det-and-loc}, Figure~\ref{fig:source-sep-MC}), detections separated by less than this threshold are considered duplicate candidates. For each such group, only the brightest source, corresponding to the highest predicted $S_1$, is retained, and the other detections are removed.

\subsection{Flux Prediction}
\label{sec:flux_pred}

Extending the original ASID framework, the present work includes a dedicated network for flux estimation. Using the $20\times20$ pixel input patches, we train an ensemble of convolutional neural networks (CNN) to predict the photon flux above 1~GeV ($S_1$), as well as its uncertainty. An ensemble of 10 identical CNNs were initialized independently with an initial learning rate of $5\times 10^{-5}$, batch size of 64 and maximum training epochs set to 300. Since $S_1$ is a continuous variable, the problem is treated as a regression task and optimized using the negative log-likelihood (NLL) loss. Owing to the broad dynamical range of the fluxes ($\sim10^{-12}$--$10^{-6}\,\mathrm{ph\,cm^{-2}\,s^{-1}}$), the network is trained on standardized values of $\log S_1$.
Model performance was tracked using validation loss and mean absolute error. For training convergence and overfitting prevention, we have also used early stopping with a patience of 20 epochs, a plateau-based learning rate decay with a factor of 0.8 and patience of 8 epochs.

The CNN architecture is based on a simplified VGG-style network \citep{VGG-16}, consisting of repeated $3\times3$ convolutional layers, max-pooling operations, and a final dense layer, without any dropout layers in between.

To improve generalization, we apply standard data augmentation techniques, including horizontal and vertical flips and rotations by $90^\circ$. To improve the representation of bright sources in the training set, we increase the number of sources above $S_1 >10^{-8}~\text{ph cm}^{-2}\text{s}^{-1}$ by $50\%$. This is a relatively modest factor, but allows us to include sources at a flux level that corresponds to the brightest sources in DR2.

\begin{figure}
    \centering
    \includegraphics[width=0.9\columnwidth]{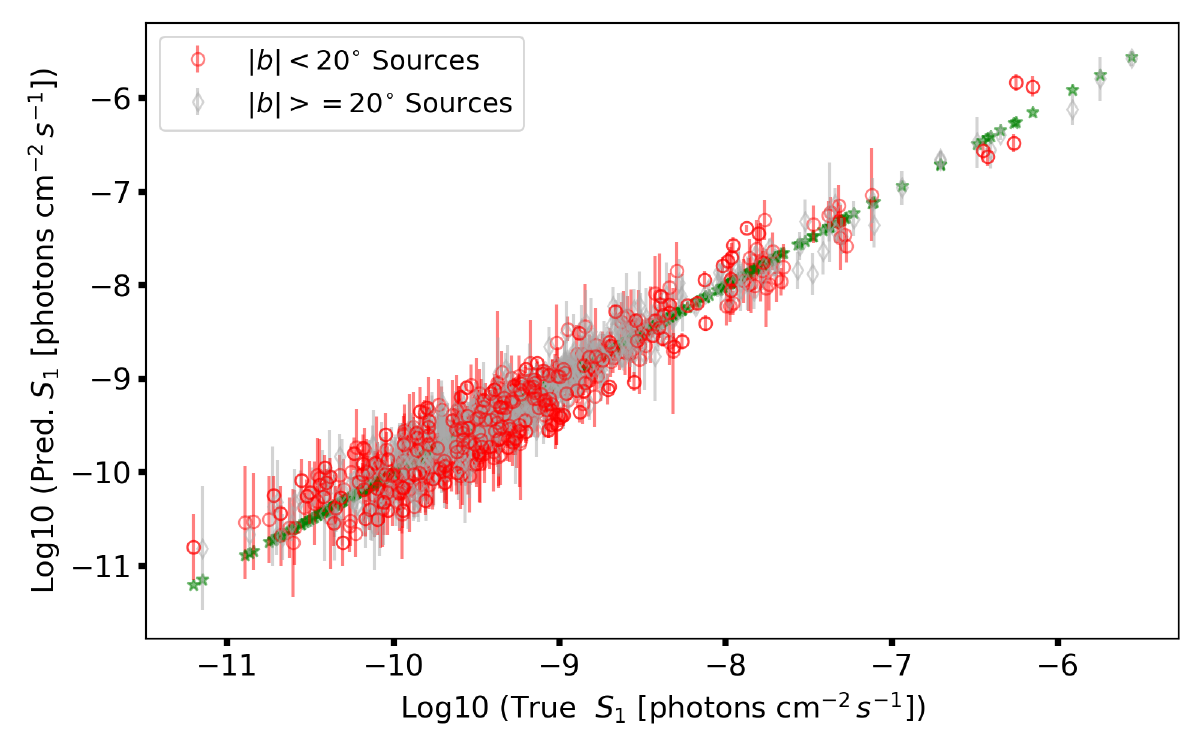}
    \includegraphics[width=0.9\columnwidth]{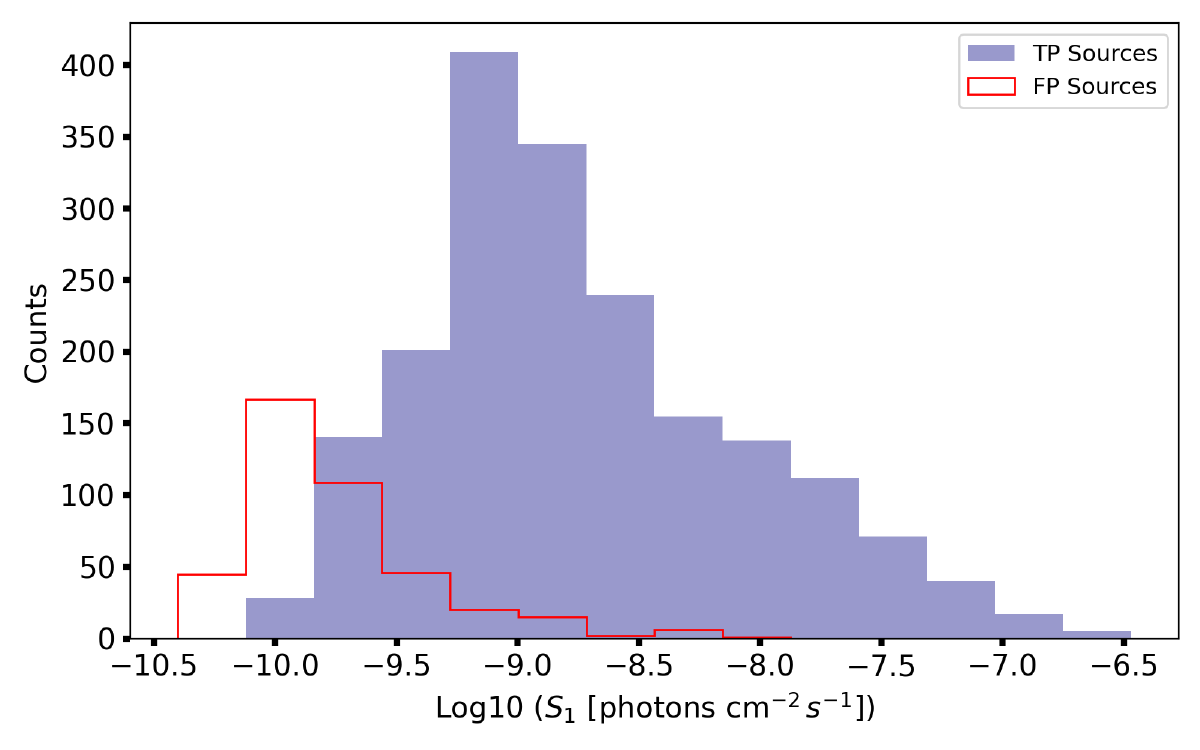}
    \caption{Top panel: Performance of flux prediction network on mock data compared with the true flux. Bottom panel: Distributions of reconstructed flux ($S_1$) for true positive (TP) and false positive (FP) detected sources.}
    \label{fig:flux-pred-train}
\end{figure}

We first validate this flux-range extension procedure using image patches centered on the true source positions (Figure~\ref{fig:flux-pred-train}). The best-performing model, selected through the minimum validation NLL loss, is then applied to all detected candidates, including FP sources. As expected, the flux predictions for high-latitude sources ($|b|>20^{\circ}$) show smaller scatter than those for low-latitude sources. We also find that FP candidates are systematically assigned lower fluxes than TP sources, as illustrated in the bottom panel of Figure~\ref{fig:flux-pred-train}.

\subsection{Classification: True vs spurious Sources}

The source detection and localization module described in Section~\ref{sec:detection} was optimized for high recall, which inevitably results in a non-negligible number of false-positive (FP) detections. In the simulated datasets, where the ground truth is known, FP candidates constitute on average $\sim18$--$19\%$ of all detections. To improve the precision while preserving high recall, we introduce a dedicated binary classifier that distinguishes true from spurious sources.

The training set is augmented using horizontal and vertical reflections as well as $90^\circ$ image rotations. The classifier architecture is similar to the CNN used for flux prediction and consists of convolutional, max-pooling, and dense layers operating on the image patches. The early stopping and learning rate reduction on plateau criteria remain the same as the flux prediction training scheme. The network outputs a score in the range $[0,1]$, where values close to 0 correspond to true sources and values close to 1 to spurious detections. A sigmoid activation function is used in the final layer, and the network is trained with the binary cross-entropy loss optimized using the Adam optimizer. To reduce overfitting, we apply early stopping if the validation loss does not improve for 10 epochs and reduce the learning rate by a factor of 0.8 after 8 stagnant epochs.

In this binary classification task, we optimized the decision threshold on the score coming from the sigmoid function to maximize the TPR while minimizing the FPR, which are defined in Eq. \ref{eq:TPR-FPR}.

\begin{equation}
\label{eq:TPR-FPR}
 \mathrm{TPR} =\frac{TP}{TP + FN},
 \qquad
 \mathrm{FPR} =\frac{FP}{TN + FP},
\end{equation}

Here, TP and FN correspond to real sources classified as true and spurious, respectively, while FP and TN denote spurious detections classified as true and spurious.

Given the inherent class imbalance in our dataset, the standard default threshold of $0.5$, which assumes an equal prior probability and symmetrical misclassification cost, was suboptimal. With this threshold, $\sim 94\%$ of the true sources were correctly classified, but at the cost of high FPR $\sim 18\%$ (i.e., $18\%$ of spurious detections were misclassified as real sources). Maximizing Youden's index ($J = \mathrm{TPR} - \mathrm{FPR}$) over the ROC curve (Receiver Operating Characteristic) yields an optimal threshold of $0.11$, which retains $\sim 89\%$ of the true sources while reducing the false positive rate on spurious detections to $\sim 7\%$.

In Appendix~\ref{sec:app-classification}, we further demonstrate that the classifier is well calibrated, with predicted probabilities consistent with the observed TP and FP rates.

\subsection{Location Refinement}

Finally, we introduce a deep ensemble regression network to refine source positions and estimate their associated uncertainties. The model takes as input $20\times 20$ pixel patches ($\sim 2^{\circ}\times2^{\circ}$) centered on the initially predicted locations from the LoG as described in Sec. \ref{sec:localization}, together with the preliminary coordinate estimates, forming a multi-input architecture. The network is trained using an NLL loss to predict refined positions and their uncertainties. The network architecture consists of a series of convolution and maxpooling layers, with dense layer at the end. The predicted variables are treated as regression variables, and thus a linear activation function is used for training with the Adam optimizer and 300 epochs. While the starting learning rate is set to $10^{-4}$, early stopping and learning rate reduction on plateau follow the same schemes as in flux or class prediction networks.

\begin{figure}
    \centering
    \includegraphics[width=0.9\columnwidth]{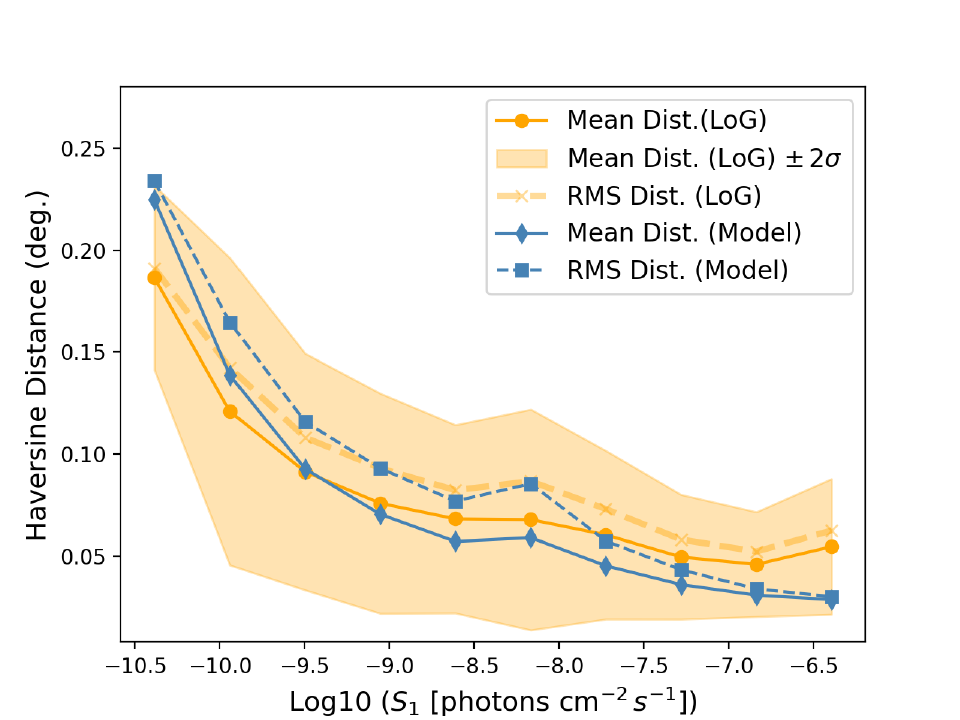}
    \caption{Mean (solid line) and RMS (dashed line) angular distances between true and predicted coordinates using the Deep Ensemble model (in blue), and the LoG (in orange) along with the $2\sigma$ band (orange patch), as a function of the logarithm of the reconstructed flux ($S_1$).
    }
    \label{fig:flux-loc-uncertain-together}
\end{figure}

To estimate epistemic uncertainty, we train 10 independent instances of the same architecture with different random weight initializations. The final prediction is obtained by averaging the outputs, while the dispersion across the ensemble provides an estimate of the positional uncertainties $(\sigma_x,\sigma_y)$ associated with the mean coordinates $(\mu_x,\mu_y)$.
The ensemble approach consistently yields smaller offsets, and provides better localization accuracy compared to LoG predictions. In Figure~\ref{fig:flux-loc-uncertain-together}, we further show the angular separation as a function of flux above 1~GeV for both methods. The ensemble performs particularly well for sources with $\log_{10} S_1 > -9.5$, corresponding to the flux range where most sources are detected (see Fig.~\ref{fig:flux-pred-train}). Overall, more than $97\%$ of true-positive sources are reconstructed within $0.2^{\circ}$ using the deep ensemble method.

    \section{Robustness to Diffuse Emission Modelling}
\label{sec:bkg-syst}

To assess the robustness of our source detection and localization pipeline, we consider two IEMs: B1, corresponding to the IEM adopted in the 4FGL catalog \citep{4fgl2020}, and B2, corresponding to the IEM used for the 3FGL catalog \citep{Acero2015:3FGL, 2016ApJS..223...26A}, rescaled to match the 4FGL IRFs. 
The 4FGL IEM has several improvements relative to the 3FGL IEM, including the use of 
Planck dust maps \citep{2016A&A...596A.109P} for the derivation of the dark neutral medium
and the new atomic hydrogen survey HI4PI \citep{2016A&A...594A.116H}, which were unavailable at the time of the creation of the 3FGL IEM.

As a first step, the pipeline is trained using simulations based on the B1 IEM and subsequently tested on data simulated with both the B1 and B2 IEMs.
In this setup, B2 acts as a proxy for an alternative (and unknown) true background, while B1 represents the model assumed during training.

\begin{figure*}
    \centering
    \begin{subfigure}[b]{0.48\textwidth}
        \centering
        \includegraphics[width=\textwidth]{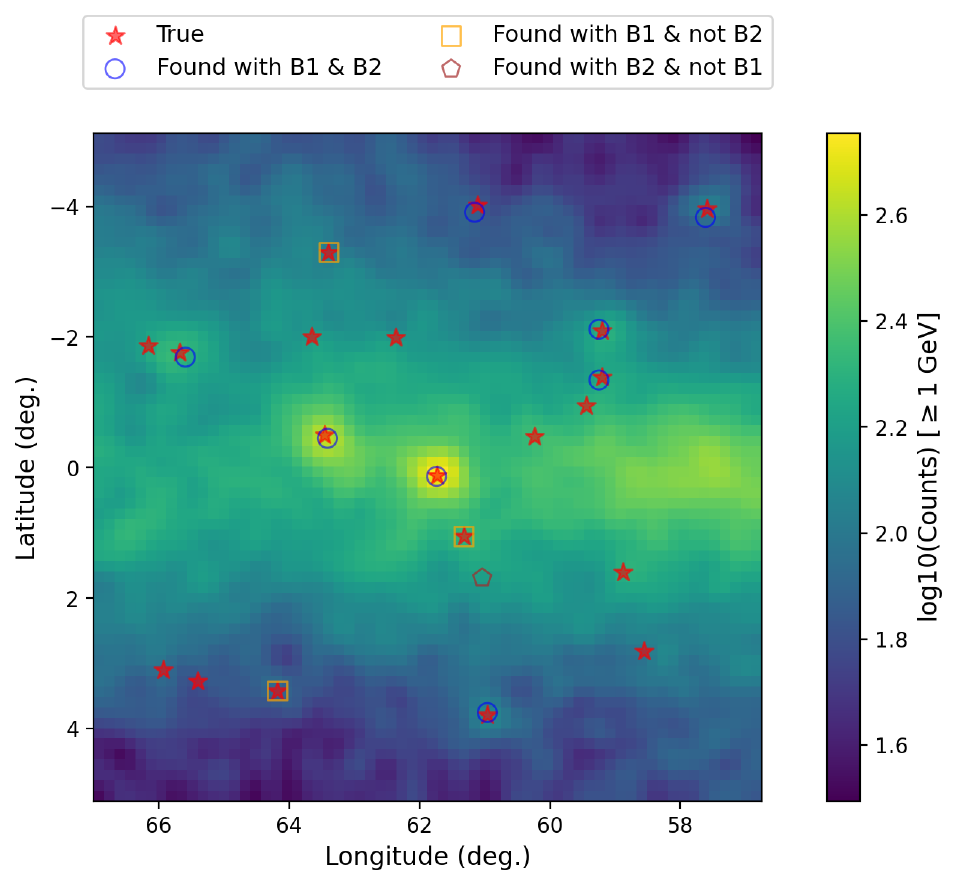}
    \end{subfigure}
    \hfill
    \begin{subfigure}[b]{0.48\textwidth}
        \centering
        \includegraphics[width=\textwidth]{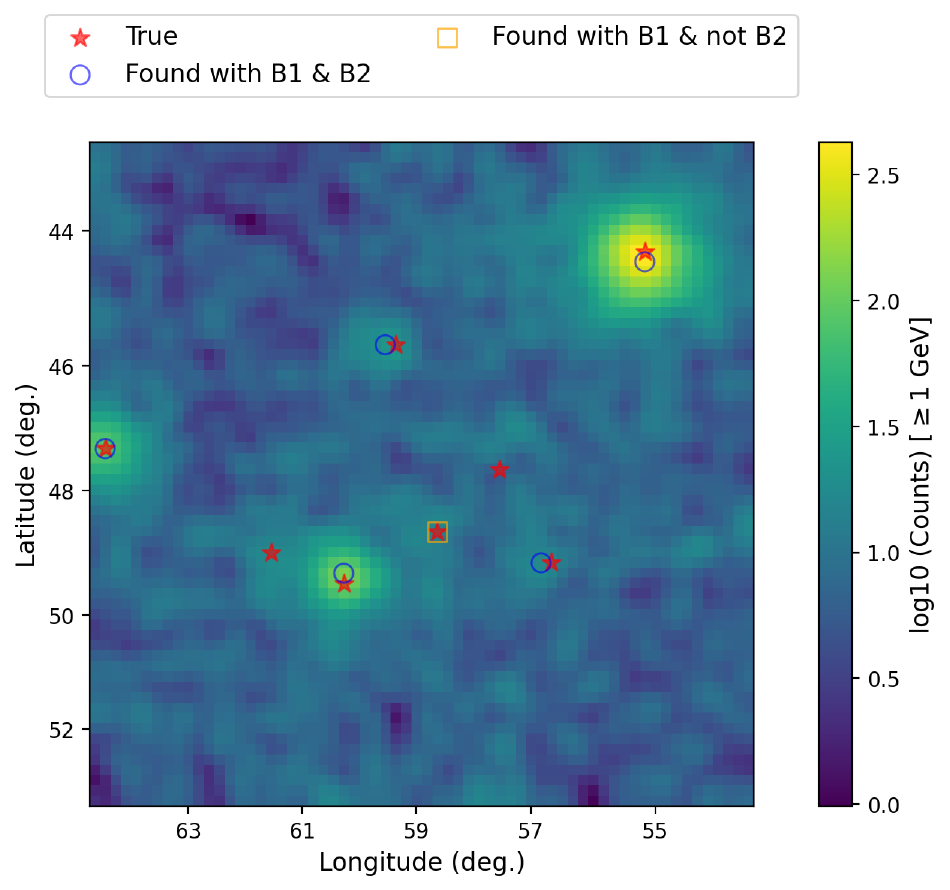}
    \end{subfigure}
    \caption{\small Example patches are shown for regions near the Galactic plane (left) and away from it (right), displaying true source positions and predicted detections using the B2 IEM. Sources detected with the B1 IEM but not with B2 are marked as orange squares, while a FP detected only with the B2 IEM is shown as an orange pentagon in the left panel. Red stars without any associated detection represent sources missed in both cases.
    }
    \label{fig:bkg-b1-b3-patch}
\end{figure*}

This comparison allows us to quantify the impact of diffuse-model mismatches on detection performance, beyond statistical fluctuations. Examples of detected sources under both background models are shown in Figure~\ref{fig:bkg-b1-b3-patch} for low ($|b|<20^{\circ}$) and high ($|b|>20^{\circ}$) latitudes. 
Orange squares, corresponding to sources detected only with B1, are examples of sources that are missed due to the change in the background model, in the examples on these plots all such sources are also true sources, so they contribute to the list of false negative sources. The purple pentagon shows a source detected with B2 background but not with B1. Since there is no true source behind the pentagon marker, this is an example of a false positive detection due to the change in the background. There are also a number of true sources, which are not detected either with B1 or with B2 backgrounds (stand-alone red stars) -- these are examples of false negative sources for both backgrounds.

A direct comparison with the standard maximum-likelihood pipeline is non-trivial, since the true source content of LAT catalogs is unknown. We therefore adopt a proxy approach based on catalog cross-comparison between FL8Y and 4FGL, which are constructed from the same data set but use different diffuse models (B2-like for FL8Y and B1 for 4FGL). Treating one catalog as reference, we define effective quantities:
$\text{TP}_{\text{eff}}$ (sources detected in both catalogs), $\text{FP}_{\text{eff}}$ (detected only in FL8Y), and $\text{FN}_{\text{eff}}$ (detected only in 4FGL). In our case, B1-based detections play the role of 4FGL, while B2-based detections correspond to FL8Y. This yields
$\text{TP}_{\text{eff}}$: common detections, $\mathrm{FP}_{\mathrm{eff}}$: B2-only detections, and $\text{FN}_{\text{eff}}$: B1-only detections (Figure~\ref{fig:bkg-b1-b3-patch}).
We then define effective precision and recall as
\[
\text{Precision}_{\text{eff}}=\frac{\text{TP}_{\text{eff}}}{\text{TP}_{\text{eff}}+\text{FP}_{\text{eff}}},
\qquad
\text{Recall}_{\text{eff}}=\frac{\text{TP}_{\text{eff}}}{\text{TP}_{\text{eff}}+\text{FN}_{\text{eff}}},
\]
which provide a practical proxy for performance in the absence of ground truth. We emphasize that these quantities are estimates and do not account for sources missed by both pipelines (red stars without counterparts in either model).

\begin{figure*}
        \centering
        \begin{subfigure}[b]{0.48\textwidth}
            \centering
            \includegraphics[width=\textwidth]{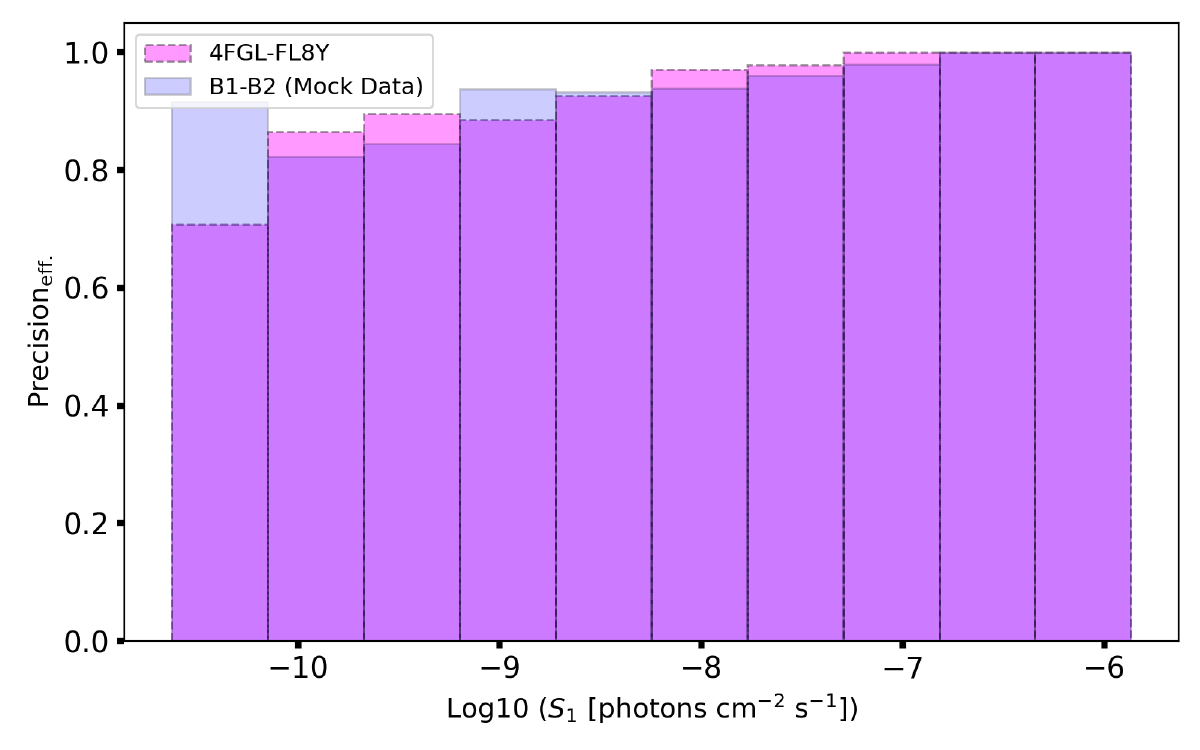}
            
        \end{subfigure}
        \hfill
        \begin{subfigure}[b]{0.48\textwidth}  
            \centering 
            \includegraphics[width=\textwidth]{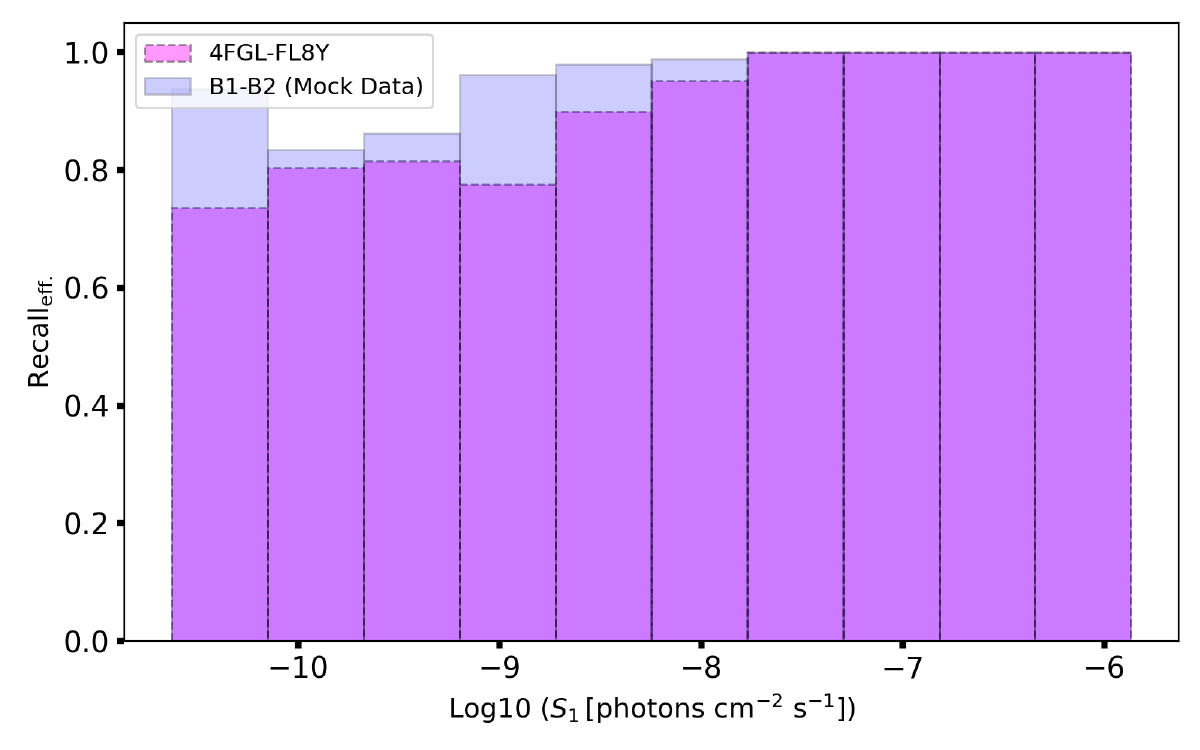}
        \end{subfigure}
        \caption{\small Left (right) panel:  $\text{Precision}_{\text{eff}}$ and ($\text{Recall}_{\text{eff}}$) rates, obtained as a function of the $S_1$ (in log \textrm{Flux1000} bins) for B1 source list, from the comparison between the two background models used in the mock data (grey bars) is shown in contrast to the $\text{Precision}_{\text{eff}}$  ($\text{Recall}_{\text{eff}}$) derived from the FL8Y–4FGL catalog comparison (magenta bars).}    
        \label{fig:bkg-sys-Pre-Rec-Eff}
    \end{figure*}

Figure~\ref{fig:bkg-sys-Pre-Rec-Eff} compares the effective precision and recall as a function of reconstructed flux above 1~GeV ($S_1$) for both FL8Y--4FGL and B1--B2 comparisons. The lowest flux bin shows artificially high effective performance (>90\%), and since the detection threshold, as discussed in subsection \ref{sec:perform-eval-detect}, is  $S_1 > 2.3\times 10^{-10}$, the effective performance should be considered from the second-to-lowest flux bin.

\begin{figure*}
        \centering
        \begin{subfigure}[b]{0.48\textwidth}
            \centering
            \includegraphics[width=\textwidth]{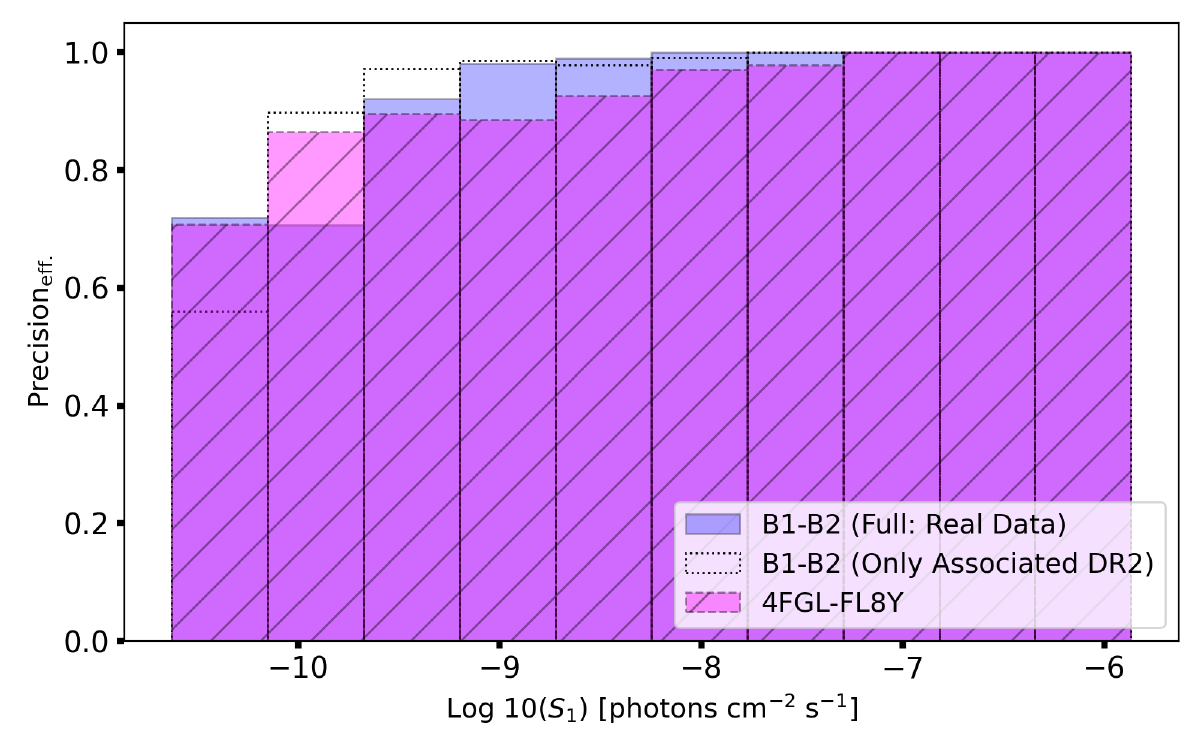}
            
        \end{subfigure}
        \hfill
        \begin{subfigure}[b]{0.48\textwidth}  
            \centering 
            \includegraphics[width=\textwidth]{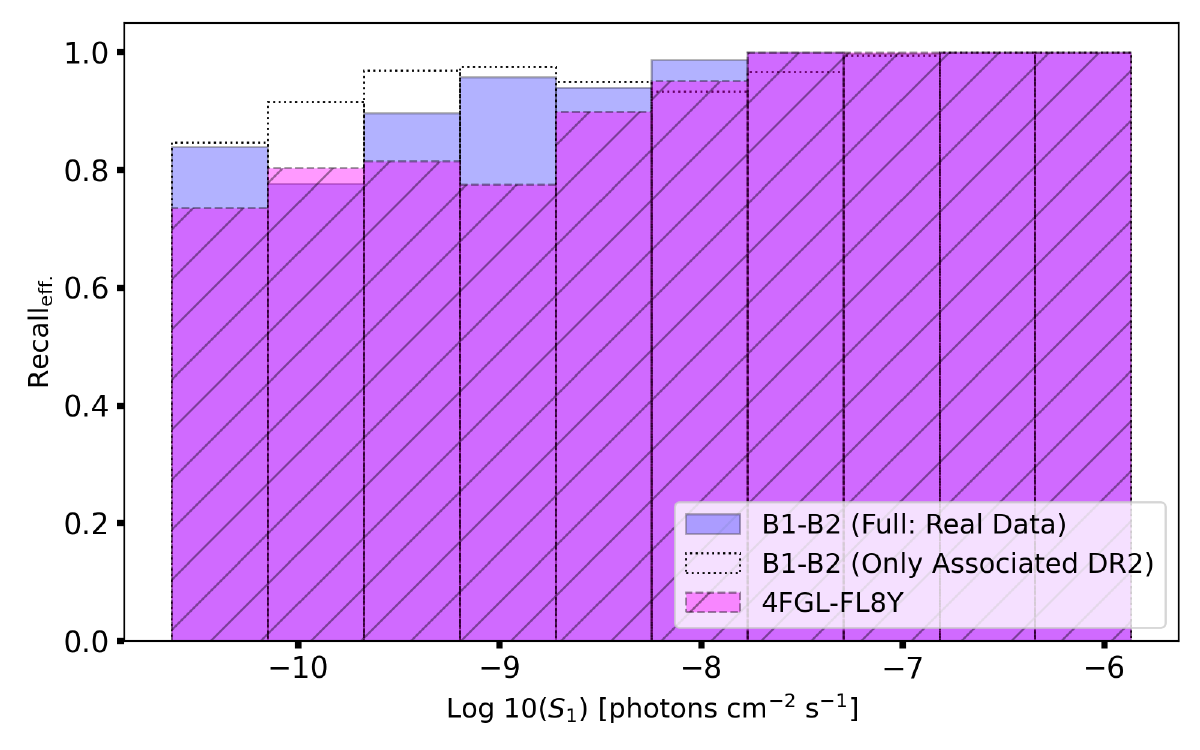}
        \end{subfigure}
        \caption{\small Same as in Figure \ref{fig:bkg-sys-Pre-Rec-Eff}, but here B1 and B2 models are applied to the real data (shown in grey bars). This is compared with the recall rate between B1 and B2 applied only to the associated DR2 sources (dotted lines). Recall between FGL-FL8Y is shown in hashed magenta bars.}    
        \label{fig:Recall-DR2-ASID-FL8Y-DR4}
    \end{figure*}

In order to have a more direct comparison of the performance of our pipeline relative to the FL8Y vs the 4FGL test, in the second step we train two models: one with B1 and one with B2 IEMs.
We then apply the two models to the real data and compare the corresponding effective precision and recall with the FL8Y vs 4FGL case in Figure \ref{fig:Recall-DR2-ASID-FL8Y-DR4}. Application of our model on the real data is described in detail in Section \ref{sec:realdata}.

We show both $\text{Precision}_{\text{eff}}$ and $\text{Recall}_{\text{eff}}$ for these two models, which are plotted with gray bars. In this case, we use $S_1$ obtained from our pipeline, which is on the x-axis. 
 In order to test the effect of the uncertainty on the flux reconstruction, we take a subset of our sources with the associations in 4FGL-DR2 and show both $\text{Precision}_{\text{eff}}$ and $\text{Recall}_{\text{eff}}$, where for this subset of sources the reported $S_1$ from 4FGL-DR2 catalog are used. 
The Figure confirms that our pipeline seems more robust to the uncertainties of the IEM, also for real data applications and independently of the flux reconstruction procedure. Later in Appendix \ref{app:flux-prediction-DR2}, we discuss the comparisons of $S_1$ predicted from our models with the 4FGL-DR2 values.

Overall, we find that our pipeline achieves comparable or better effective precision and recall than the likelihood-based catalog pipeline with respect to changing IEMs.

\section{1FDL Source List}
\subsection{Source List Creation}
\label{sec:realdata}

In this section, we present the application of our pipeline to 10 years of {\Fermi}-LAT observations and the construction of the first deep-learning-based $\gamma$-ray source list, 1FDL. As for the simulated data, the pipeline provides source positions, flux estimates, and probabilities for sources to be genuine or spurious.

The list construction proceeds through the following steps:
\begin{itemize}
    \item generation of segmented masks using the trained MIUN together with source coordinates predicted by the LoG algorithm;
    \item extraction of $20\times20$ pixel patches centered on the predicted source positions;
    \item refinement of source locations and estimation of positional uncertainties;
    \item classification of candidates into true and spurious sources;
    \item estimation of the photon flux above 1~GeV using the flux-prediction network;
    \item creation of the final source table containing positions, fluxes, and classification scores.
\end{itemize}

The initial detection and localization stage yields more than 6000 source candidates. Since the recovery efficiency for close source pairs decreases significantly below angular separations of $0.1^{\circ}$ (Appendix~\ref{app:det-and-loc}, Figure~\ref{fig:source-sep-MC}), multiple detections within this radius are treated as duplicate candidates. In such cases, only the brightest source (highest predicted $S_1$) is retained, while the remaining detections are discarded.

After this cleaning step, the list contains more than 5100 candidates. Applying the true--spurious classifier further removes low-confidence detections, resulting in a clean sample of 3821 sources. Starting from this list, we search for associations with the 4FGL-DR2 catalog, which contains more than 5700 sources. The 1FDL source list is publicly available at [GitHub + Zenodo in preparation].

The association procedure with 4FGL-DR2 is performed as follows:
\begin{itemize}
    \item both the 1FDL and 4FGL-DR2 source lists are ordered by decreasing flux ($S_1$);
    \item associations are determined by iterating over the 1FDL sources and searching for counterparts in the DR2 list. We verified that the results are unchanged when starting instead from the DR2 catalog;
    \item a matching radius of $0.25^{\circ}$ is adopted, consistent with the criterion used for the simulated data;
    \item once a 1FDL source is associated with a DR2 source, that DR2 source is stored as its association and removed so that further associations are unique. If no DR2 association is found, the algorithm likewise proceeds to the next 1FDL source.
\end{itemize}

\subsection{1FDL Source List Summary}
The 1FDL contains 3821 sources detected over the whole sky using 10 years of \textit{Fermi}-LAT data. A full sky-map of these sources is shown in Figure \ref{fig:1FDL-allSky}. 
A summary of the associated sources and their corresponding 4FGL-DR2 classes (\textrm{CLASS1}) is provided in Table~\ref{tab:assoc-class}. As expected, a large fraction of the associations correspond to extragalactic sources, primarily BLLs, FSRQs, and BCUs. A substantial number of associated sources are unclassified in 4FGL-DR2, consistent with the fact that nearly $\sim28\%$ of DR2 sources remain unassociated. 
A full description of the 1FDL columns is provided in Table~\ref{tab:1fdl-cols}.

Following the introduction of the Soft Galactic Unassociated (SGU) class in 4FGL-DR3 \citep{4FGL-DR3}, defined as unassociated inner-Galaxy sources ($|b|<10^{\circ}$) with soft spectra (\textrm{PL\_Index}>2.4), we additionally define a Hard Galactic Unassociated (HGU) class with \textrm{PL\_Index}<2.4. Within the 4FGL-DR2 catalog, we identify 583 SGUs and 246 HGUs, of which $\sim19\%$ and $\sim28\%$, respectively, are associated with 1FDL.

\begin{table*}
    \caption{Source classes of 1FDL sources after association with the 4FGL-DR2 catalog. The designator column corresponds to the  \textrm{`CLASS1'} field in the 4FGL-DR2. `Unassociated in DR2' denotes DR2 sources without counterparts at other wavelengths, while `Other Galactic' and `Other Extragalactic' group the remaining Galactic and extragalactic DR2 source classes, respectively. `No DR2 Counterpart' refers to 1FDL sources without an associated source in the 4FGL-DR2 catalog.}
    \centering
    \begin{tabular}{ccc}
    \hline \hline
        Description &  Designator & Number\\
        \\
        \hline
        \\
        BL Lac type of blazar & bll + BLL & 954 + 22 \\
        \\
        FSRQ type of blazar & fsrq + FSRQ & 496 + 43 \\
        \\
        Blazar candidate of uncertain type & bcu + BCU & 629 + 2 \\
        \\
        Radio Galaxy & rdg + RDG & 17 + 4 \\
        \\
        Pulsar & psr + PSR & 21 + 200 \\
        \\
        Pulsar Wind Nebula & pwn + PWN & 2 + 3\\
        \\
        Supernova Remnant & snr + SNR & 9 + 12\\
        \\
        Supernova Remnant/Pulsar Wind Nebula & spp + SPP & 31 + 0 \\
        \\
        Unknown & unk + UNK & 16 + 0 \\
        \\
        Other Galactic & --- &  30 \\
        \\
        Other Extragalactic & --- &  32 \\
        \\
        Unassociated in DR2 & DR2 counterpart is unassociated & 482 \\
        \\
        No DR2 Counterpart & --- & 816 \\ 
        \hline
        Total 1DFL & --- & 3821\\
        \hline 
    \end{tabular}
    
    \label{tab:assoc-class}
\end{table*}

\begin{figure*}
        \centering
        \begin{subfigure}[b]{0.49\textwidth}
            \centering
            \includegraphics[width=\textwidth]{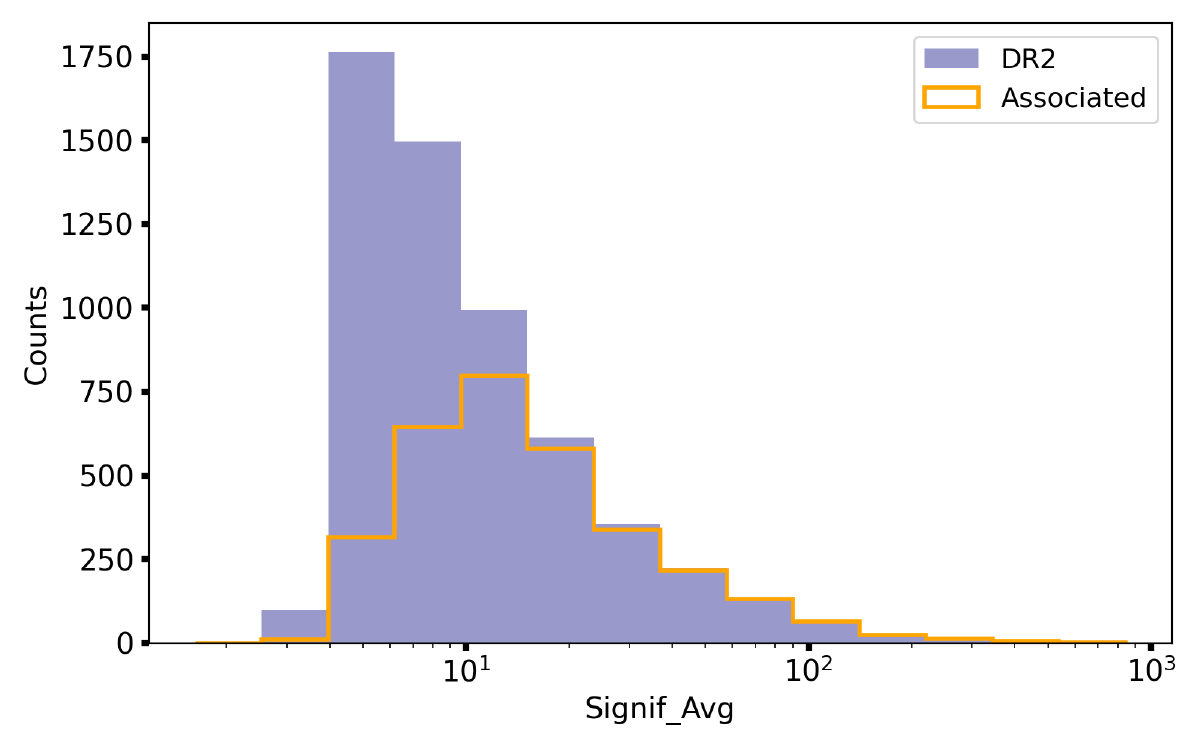}
        \end{subfigure}
        \hfill
        \begin{subfigure}[b]{0.49\textwidth}  
            \centering 
            \includegraphics[width=\textwidth]{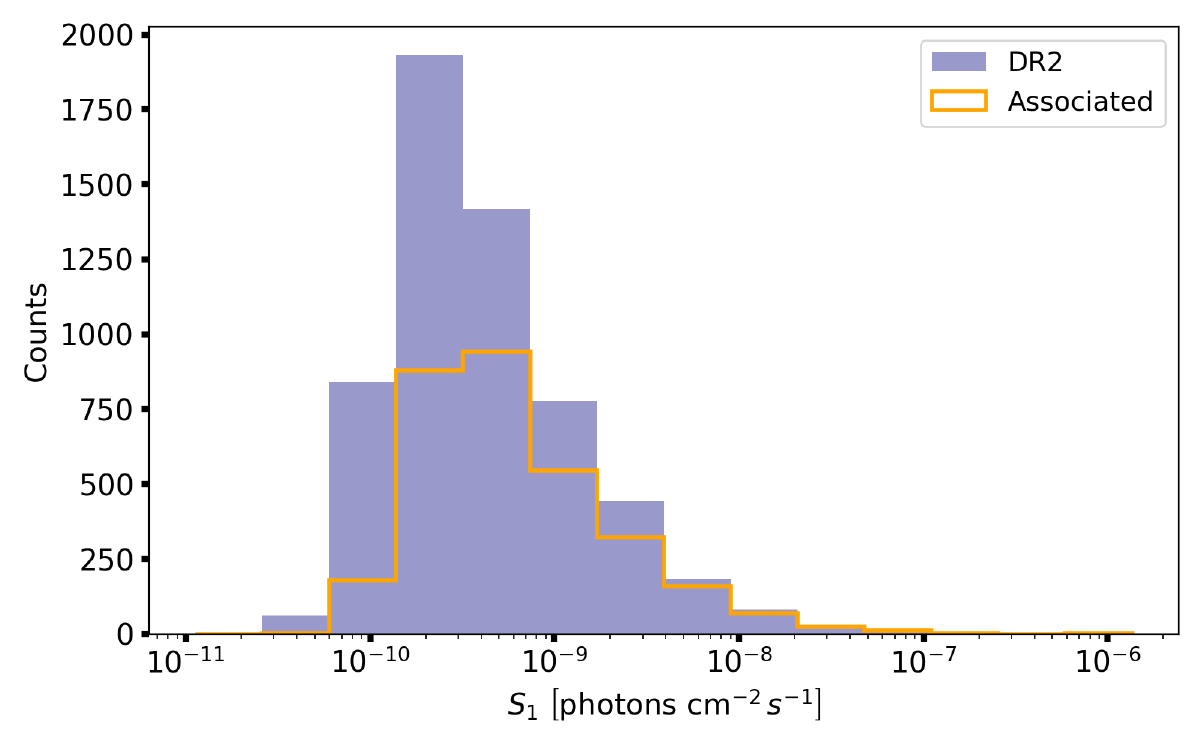}
        \end{subfigure}
        \caption{\small Left: Number of 1FDL associations with 4FGL-DR2 sources as a function of the DR2 detection significance, \textrm{Signif\_Avg}, over the full sky. Right: Number of associations as a function of the DR2 photon flux, \textrm{Flux1000}, over the full sky.} 
        \label{fig:assoc-rate-DR2-sigma}
    \end{figure*}

\begin{figure*}
    \centering
    \includegraphics[width=0.9\linewidth]{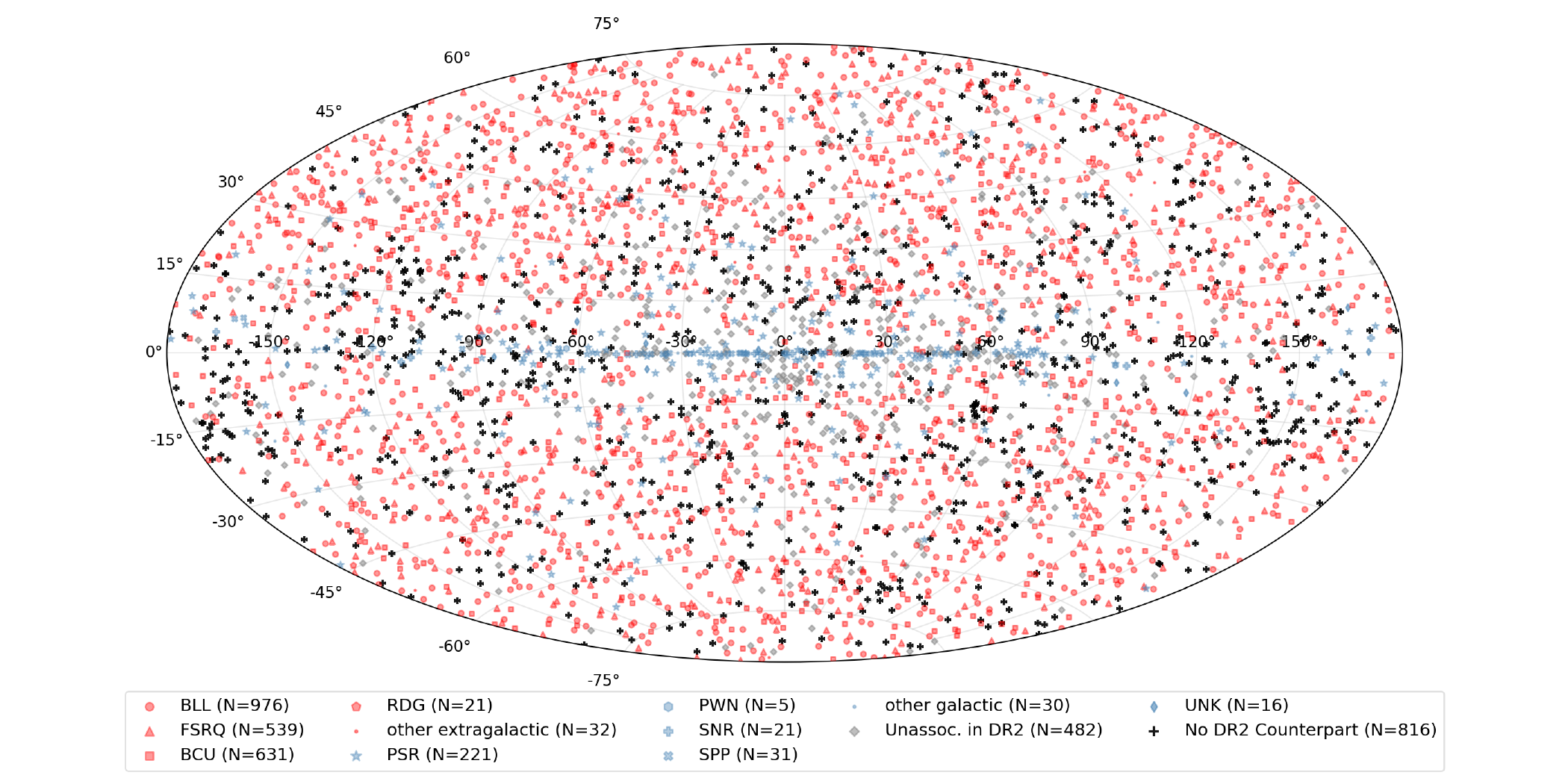}
    \caption{Full sky map showing all 1FDL sources as listed in Table \ref{tab:assoc-class}. We have used the DR2 source classes (`CLASS1') to label the already associated sources from the 1FDL list to the 4FGL-DR2 catalog.  1FDL sources with no counterpart in DR2 catalog are shown with `$+$' marker. Other source classes and corresponding association numbers follow as described in Table \ref{tab:assoc-class}.}
    \label{fig:1FDL-allSky}
\end{figure*}

We further investigate the 1FDL source properties and compare them with those of the 4FGL-DR2 catalog. We first examine the fraction of 4FGL-DR2 sources successfully associated with 1FDL as a function of their \textrm{Signif\_Avg} values. Table~\ref{tab:Assoc-DR2} summarizes the results and shows that our pipeline achieves an association rate $\sim 75\%$ for all 4FGL-DR2 sources with \textrm{Signif\_Avg}$>10$ across the full sky, increasing to $\sim 93\%$ for bright sources with $\textrm{Signif\_Avg}>20$.
Figure~\ref{fig:assoc-rate-DR2-sigma} presents the total number of 4FGL-DR2 sources together with the subset associated with 1FDL as a function of $\textrm{Signif\_Avg}$ and $\textrm{Flux1000}$. The association rate decreases for faint sources with $\textrm{Signif\_Avg}<10$ and $\textrm{Flux1000}<10^{-9}\,\mathrm{ph~ cm^{-2} ~s^{-1}}$. We note that, within the 4FGL-DR2 catalog itself, the fraction of sources associated at other wavelengths also declines substantially in this regime, from $\sim 75\%$ for high-latitude sources $(|b|>20^{\circ})$ to $\sim 44\%$ near the Galactic plane $(|b|<20^{\circ})$. Since no directly comparable deep-learning pipeline operating on full-sky $\gamma$-ray photon data currently exists, these results provide a standalone benchmark. Note that \cite{2025PhRvD.112j3022E} recently reported a recovery rate of $\sim 98\%$ for bright, unflagged 4FGL-DR4 high-latitude sources $(|b|>30^{\circ})$ with $S_1 > 3\times10^{-10}~\mathrm{ph~cm^{-2}~s^{-1}}$ using a neural ratio estimation method. Under equivalent conditions (bright, unflagged, and extragalactic sources), our pipeline also recovers $\sim 98\%$ of the 4FGL-DR2 sources.

\begin{figure}
    \centering
    \includegraphics[width=0.9\linewidth]{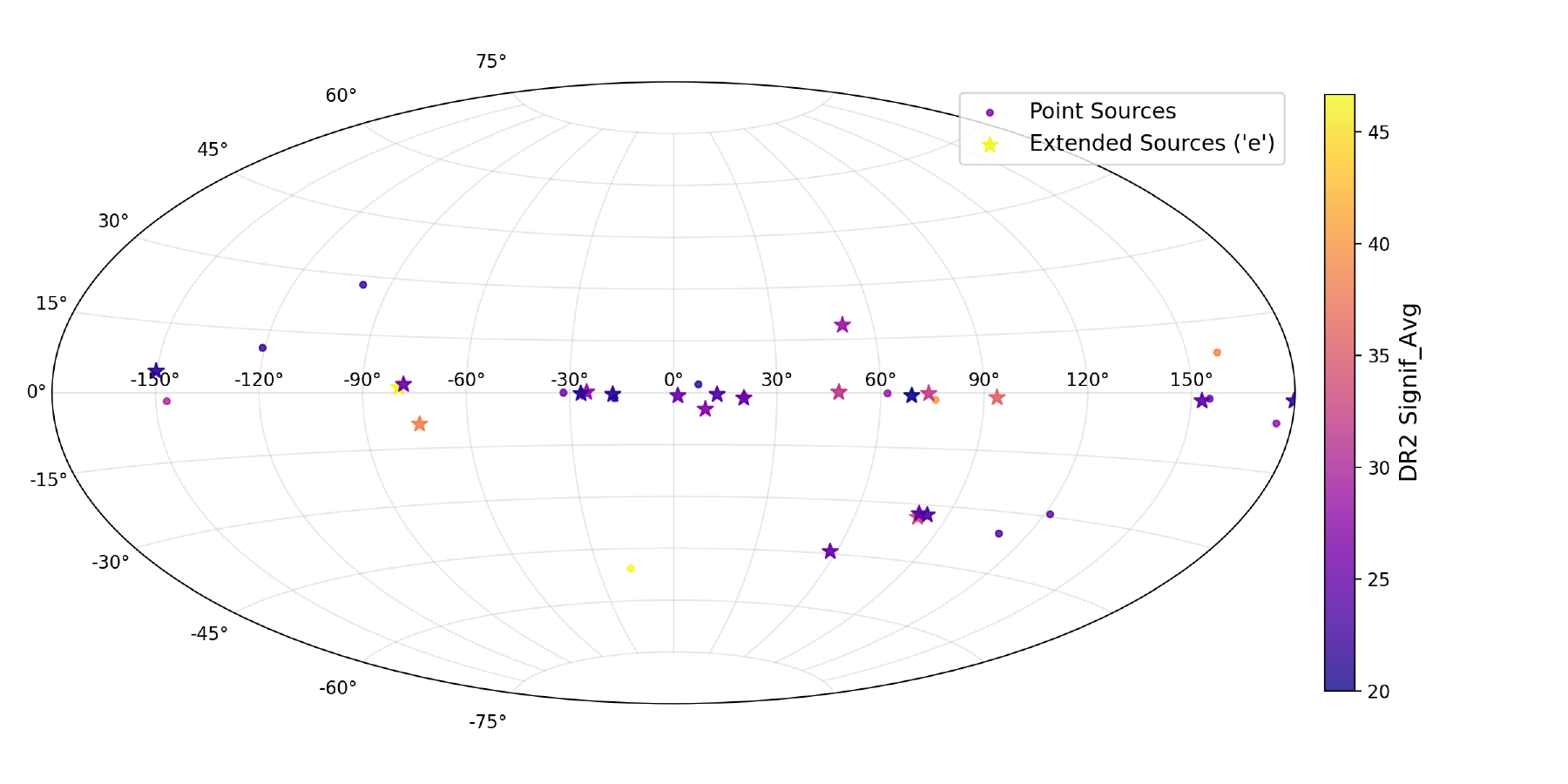}
    \caption{4FGL-DR2 sources with \textrm{Signif\_Avg} $>20$ that are not associated with 1FDL. This illustrates their concentration toward the Galactic plane. Extended (`e') DR2 sources are marked with stars.}
    \label{fig:DR2-Unassoc-HSignif}
\end{figure}

We further find that high-significance DR2 sources without 1FDL counterparts are predominantly concentrated near the Galactic plane, as illustrated in Figure~\ref{fig:DR2-Unassoc-HSignif}. In the same figure, the 22 extended 4FGL-DR2 sources (`e') are marked with stars. 
The absence of associations for some bright sources likely reflects limitations of the current pipeline, which was optimized exclusively for point-source detection and trained only on simulated point-like emitters. A specific source-by-source inspection leads to the cause of unassociation for these sources: i) either from possible larger uncertainty in localization, where the network predicted location is offset from DR2 positions by more than the association radius of $0.25^{\circ}$ (as shown in the top panel of Figure \ref{fig:unassoc-bright-2patch}); ii) or non-detection of sources in bright, crowded fields where the source confusion could be caused by high local source density and/or bright IEM as shown in the bottom panel of Figure \ref{fig:unassoc-bright-2patch}.

Remaining non-extended bright unassociated DR2 sources have soft spectra $(\textrm{PL\_Index}>2.4)$, whose emission is dominated by photons below $\sim 2$~GeV, where the LAT angular resolution is significantly poorer. Consequently, the dominant limitation is source confusion and overlapping emission, rather than insufficient source brightness alone.

In total, 816 1FDL sources have no counterpart in the 4FGL-DR2 catalogue (Table~\ref{tab:assoc-class}), and may therefore represent previously unidentified gamma-ray sources and/or residuals from imperfect modeling of the Galactic diffuse emission. Most of these sources are either faint, located close to the Galactic plane, or lie in crowded regions near already associated 4FGL-DR2 and 1FDL sources. However, ten sources are located outside the Galactic plane, have relatively bright predicted fluxes $\left(S_1 > 5\times10^{-10},\mathrm{ph,cm^{-2},s^{-1}}\right)$, and are separated by more than $0.25^{\circ}$ from both associated 4FGL-DR2 sources and other 1FDL sources, standing out as particularly promising new-source candidates. A closer inspection, however, indicates that seven of these ten candidates (1FDL\,J0329.0+3119, 1FDL\,J0333.5+3105, 1FDL\,J1553.9-0448, 1FDL\,J1554.4-0256, 1FDL\,J1621.6-2012, 1FDL\,J2052.4+7744, and 1FDL\,J2057.1+7728) are located in regions containing dense atomic or molecular gas, i.e., they may be spurious sources due to diffuse gamma-ray emission.
Another candidate, 1FDL\,J1506.2+1017, lies within $\sim0.5^{\circ}$ of the very bright $\gamma$-ray source PKS\,1502+106, raising the possibility that its detection is related to residual emission from this nearby source.
Thus, environmental and source-confusion effects provide plausible explanations for eight of the ten brightest unassociated 1FDL sources. Nevertheless, two sources, 1FDL\,J1647.9-1209 and 1FDL\,J0535.7-6610, have subsequently been confirmed as gamma-ray sources in the latest list of \textit{Fermi}-LAT sources based on 16 years of data \citep[FL16Y,][]{2026arXiv260222148B}. The confirmation of these two sources demonstrates that the 1FDL analysis can identify gamma-ray sources that were not present in 4FGL-DR2, highlighting the potential of our DNN method to uncover previously undetected gamma-ray sources.

\begin{table}
\caption{Association rate with 4FGL-DR2 sources as a function of Galactic latitude (\textrm{GLAT}) and detection significance (\textrm{Signif\_Avg}). The results are summarized in six regions defined by high ($\sigma > 20$), medium ($10 < \sigma < 20$), and low ($\sigma < 10$) significance bins at high ($|b| > 20^{\circ}$) and low ($|b| < 20^{\circ}$) Galactic latitudes. The reported rates are normalized to the total number of DR2 sources in each corresponding bin.}\label{tab:Assoc-DR2}
\centering
\begin{tabular}{cc|c|c|}
\cline{3-4}
& & \multicolumn{2}{c|}{\textrm{GLAT}} \\
\cline{3-4}
& & $|b|>20^{\circ}$ & $|b|<20^{\circ}$ \\
\hline
\multicolumn{1}{|c|}{\multirow{3}{*}{\textrm{Signif\_Avg}}} 
& $\sigma>20$ & 99\% & 93\% \\ 
\cline{2-4}
\multicolumn{1}{|c|}{} 
& $10<\sigma<20$ & 91\% & 73\% \\
\cline{2-4}
\multicolumn{1}{|c|}{} 
& $\sigma<10$ & 35\% & 19\% \\
\hline
\end{tabular}
\end{table}

\section{Conclusions and Outlook}
\label{sec:conclusions}

In this work, we present the first fully end-to-end deep-learning-based list of $\gamma$-ray sources, the 1FDL, constructed from ten years of data collected by \textit{Fermi}-LAT. The pipeline is trained on realistic simulations generated with \texttt{Fermitools}, incorporating energy-dependent instrumental effects and class-specific source properties. It performs detection, localization, and characterization in a unified framework, delivering flux estimates and a probabilistic assessment of source reliability. The 1FDL source list, which contains more than 3800 sources, is publicly available (GitHub repository link), enabling reproducibility and further community development.
Our main results can be summarized as follows:
\begin{itemize}
    \item First end-to-end DNN source list: we demonstrate, for the first time, that a fully supervised deep-learning pipeline can produce a $\gamma$-ray source list and their properties, directly from full-sky photon data, without relying on traditional likelihood methods.
    \item Robust performance against diffuse backgrounds: evaluated with two independent interstellar emission models (IEMs), the pipeline achieves effective precision and recall comparable to or exceeding those of established likelihood-based catalogs such as derived in the 4FGL catalog and FL8Y list.
    \item High-latitude performance: for $|b|>20^\circ$, both $\text{Precision}_{\text{eff}}$ and $\text{Recall}_{\text{eff}}$ remain above 90\% above the LAT detection threshold of $2\times10^{-10}~ \text{ph cm}^{-2} \text{s}^{-1}$.
    \item Recovery of known sources: when applied to ten years of data, the 1FDL recovers $>98\%$ of high-latitude sources with $\texttt{Signif\_Avg}>20$ from 4FGL-DR2, and maintains $>75\%$ completeness for $\texttt{Signif\_Avg}>10$ sources over the full sky.
    \item Probabilistic source validation: the assignment of a “real vs. spurious” probability provides a transparent and flexible quality metric that can be tuned to scientific priorities (completeness vs. purity).
    \item Of the $\sim3800$ sources contained in the 1FDL, $\sim800$ have no association in 4FGL-DR2. Among the brightest and high-latitude unassociated sources, two have subsequently been confirmed as gamma-ray sources in the {\Fermi}-LAT 16-year source list (FL16Y), demonstrating the potential of the method to uncover previously unidentified sources.
\end{itemize}

Beyond its immediate results, this work establishes a new paradigm for $\gamma$-ray catalog construction. Our results demonstrate that the deep-learning framework is complementary to likelihood-based approaches and offers a scalable alternative particularly suited to large data volumes and complex backgrounds. Moreover, expanding our studies through retraining with additional years of Fermi-LAT observations will enable continuous source list updates, while extensions to lower energies (<300 MeV) can exploit DNN capabilities in discriminating sources from the bright and extended background \citep{fle-giacomo}.
Looking forward, the methodology is directly transferable to other $\gamma$-ray surveys. It can be adapted to the data products of $\gamma$-ray ground-based instruments, 
as the next-generation Cherenkov Telescope Array Observatory \citep[CTAO,][]{science-cta}, whose unprecedented sensitivity and angular resolution will require highly automated, robust catalog-building strategies, an approach whose feasibility we have already begun to explore in preliminary studies based on simulated CTAO data \citep{judit-ctao_2025}. More broadly, the end-to-end architecture is experiment-agnostic: with appropriate instrument response modeling, it can be retrained for forthcoming $\gamma$-ray missions and surveys, as well as integrated into multi-wavelength workflows. In particular, its prior adaptation to optical catalog-building tasks \citep{Fio-AsidL, Fio-AsidC} demonstrates its versatility and points toward unified, data-driven source detection and characterization across the electromagnetic spectrum. As data volumes and instrumental complexity continue to grow, such deep-learning-driven pipelines are expected to become a central component of next-generation astrophysical surveys.

\begin{acknowledgements}

The work of S.B. and G.Z. was supported by the Slovenian Research Agency under grants P1-0031, N1-0344, and J1-60014.

The work of D.M. is supported by the DFG grant MA 8279/3-1.
G.P. acknowledges ISCRA for awarding this project access to the LEONARDO supercomputer, owned by the EuroHPC Joint Undertaking, hosted by CINECA (Italy).
\\
The authors would like to thank Benoit Lott, Jean Ballet, and Matthew Kerr for helpful discussions and comments about the work and the manuscript.
\\
The authors acknowledge the use of several software packages: \href{https://healpy.readthedocs.io/en/latest}{Healpy}~\citep{2005ApJ...622..759G, healpy},
\href{http://www.astropy.org}{Astropy}~\citep{2013A&A...558A..33A},
\href{https://matplotlib.org/}{Matplotlib}~\citep{Hunter:2007}, 
\href{https://pandas.pydata.org/}{pandas}~\citep{mckinney-proc-scipy-2010},
\href{https://scikit-learn.org/stable/}{scikit-learn}~\citep{scikit-learn},
\href{https://fermi.gsfc.nasa.gov/ssc/data/analysis/software/}{Fermitools}, \href{https://www.tensorflow.org/}{TensorFlow}~\citep{tensorflow2015-whitepaper}.
\\
The \textit{Fermi} LAT Collaboration acknowledges generous ongoing support from a number of agencies and institutes that have supported both the development and the operation of the LAT as well as scientific data analysis. These include the National Aeronautics and Space Administration and the Department of Energy in the United States, the Commissariat à l’Energie Atomique and the Centre National de la Recherche Scientifique / Institut National de Physique Nucléaire et de Physique des Particules in France, the Agenzia Spaziale Italiana and the Istituto Nazionale di Fisica Nucleare in Italy, the Ministry of Education, Culture, Sports, Science and Technology (MEXT), High Energy Accelerator Research Organization (KEK) and Japan Aerospace Exploration Agency (JAXA) in Japan, and the K. A.Wallenberg Foundation, the Swedish Research Council and the Swedish National Space Board in Sweden. Additional support for science analysis during the operations phase is gratefully acknowledged from the Istituto Nazionale di Astrofisica in Italy and the Centre National d’Etudes Spatiales in France. This work is performed in part under DOE Contract DE-AC02-76SF00515.
\end{acknowledgements}

\begin{appendix}
\section{Data Preparation}

The simulated data samples described in Section~\ref{sec:train-data} were generated using the LP and PLEC spectral parameterizations introduced in Eqs.~(\ref{eq:logparabola}) and (\ref{eq:plec}), respectively. In this appendix, we compare the distributions of the spectral parameters adopted in the simulations with those measured for the corresponding source populations in the 4FGL-DR2 catalog.

\begin{figure*}
    \centering
    \includegraphics[width=0.8\linewidth]{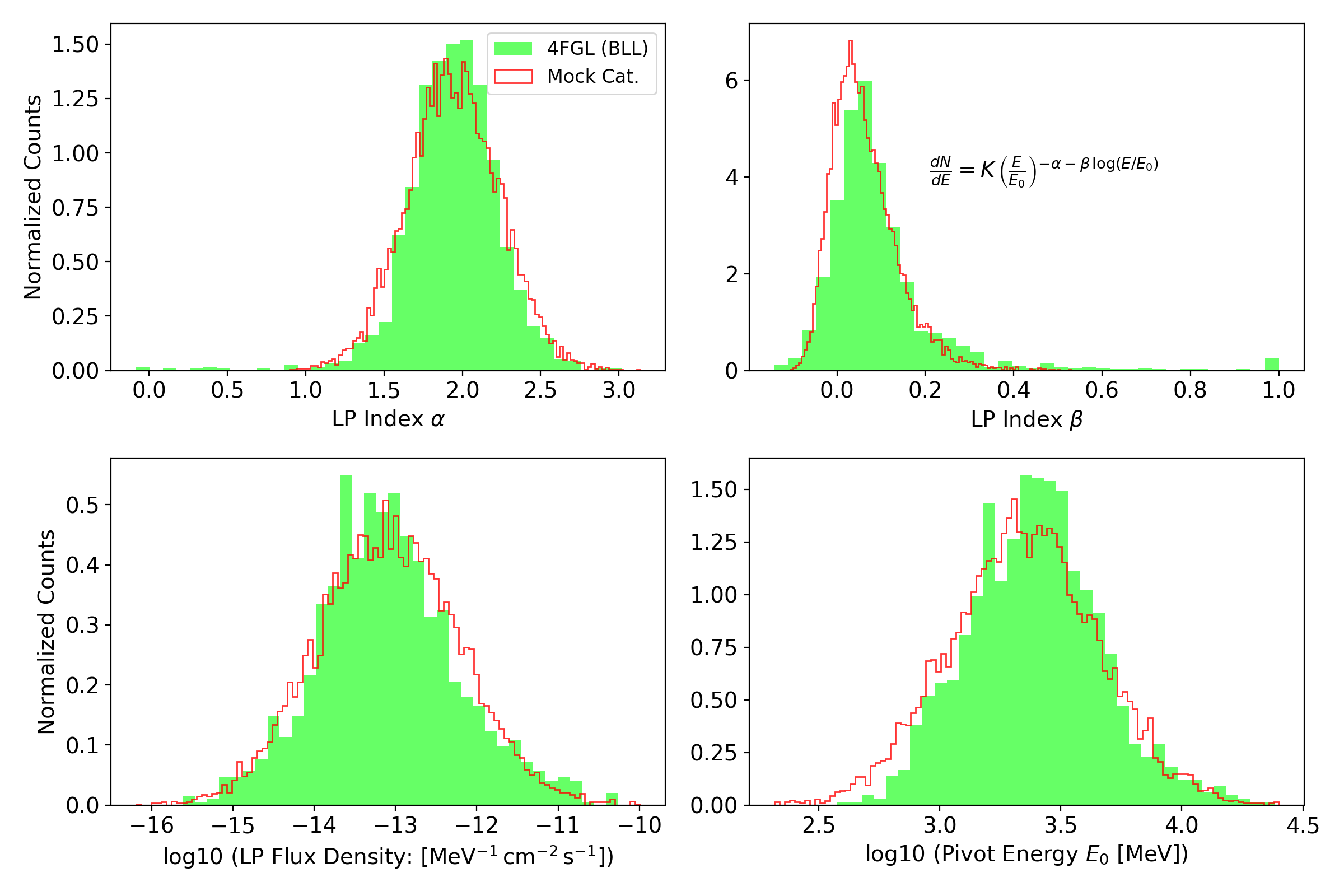}
    \caption{\small{Distributions of the LP spectral parameters for BL Lac sources. Green shaded histograms show the distributions for the 4FGL-DR2 catalog, while the red curves correspond to the simulated BL Lac population used to generate the training dataset.}
    }
    \label{fig:bll-params}
\end{figure*}

\begin{figure*}
    \centering
    \includegraphics[width=0.8\linewidth]{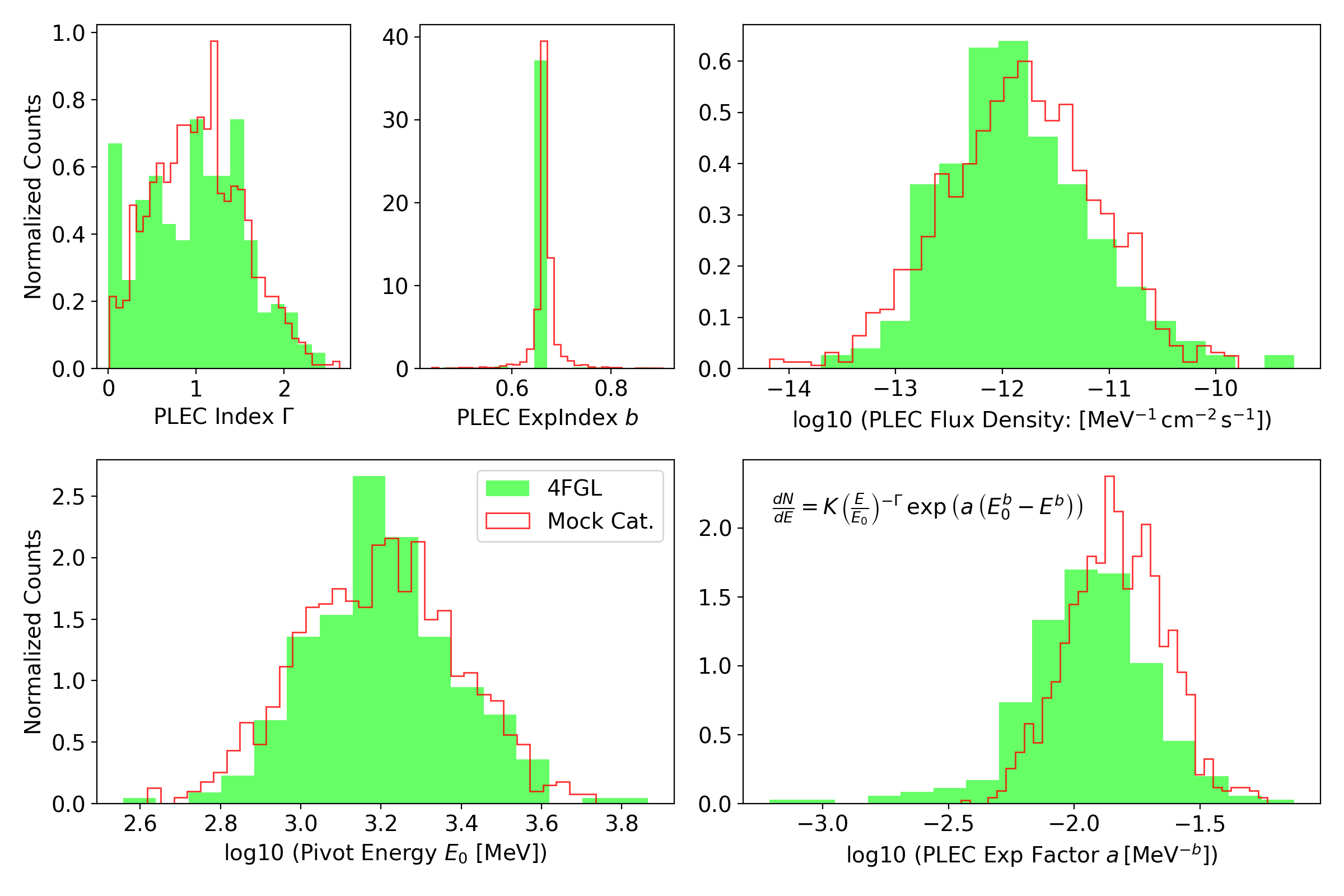}
    \caption{\small{Distributions of the PLEC spectral parameters for PSR sources. Green shaded histograms show the distributions in the 4FGL-DR2 catalog, while the red curves correspond to the simulated PSR population used in the training dataset.}
    }
    \label{fig:plec-params}
\end{figure*}

Figure~\ref{fig:bll-params} compares the distributions of the LP parameters (Eq.~\ref{eq:logparabola}) in the mock catalog and in the 4FGL-DR2 catalog, using BL Lac sources as an illustrative example. The corresponding parameter distributions for the PLEC model (Eq.~\ref{eq:plec}) are shown for the PSR population in Figure~\ref{fig:plec-params}. For the PLEC exponential index b, we used the observed standard deviation 0.02, although the standard deviation obtained only from the 7 pulsars in which that parameter was free would be 0.13.

The spatial distributions of BLL and FSRQ sources were assumed to be isotropic. For the PWN-SNR population, the latitude and longitude distributions were modeled with Gaussian functions. 

\begin{figure}
    \centering
    \includegraphics[width=0.96\columnwidth]{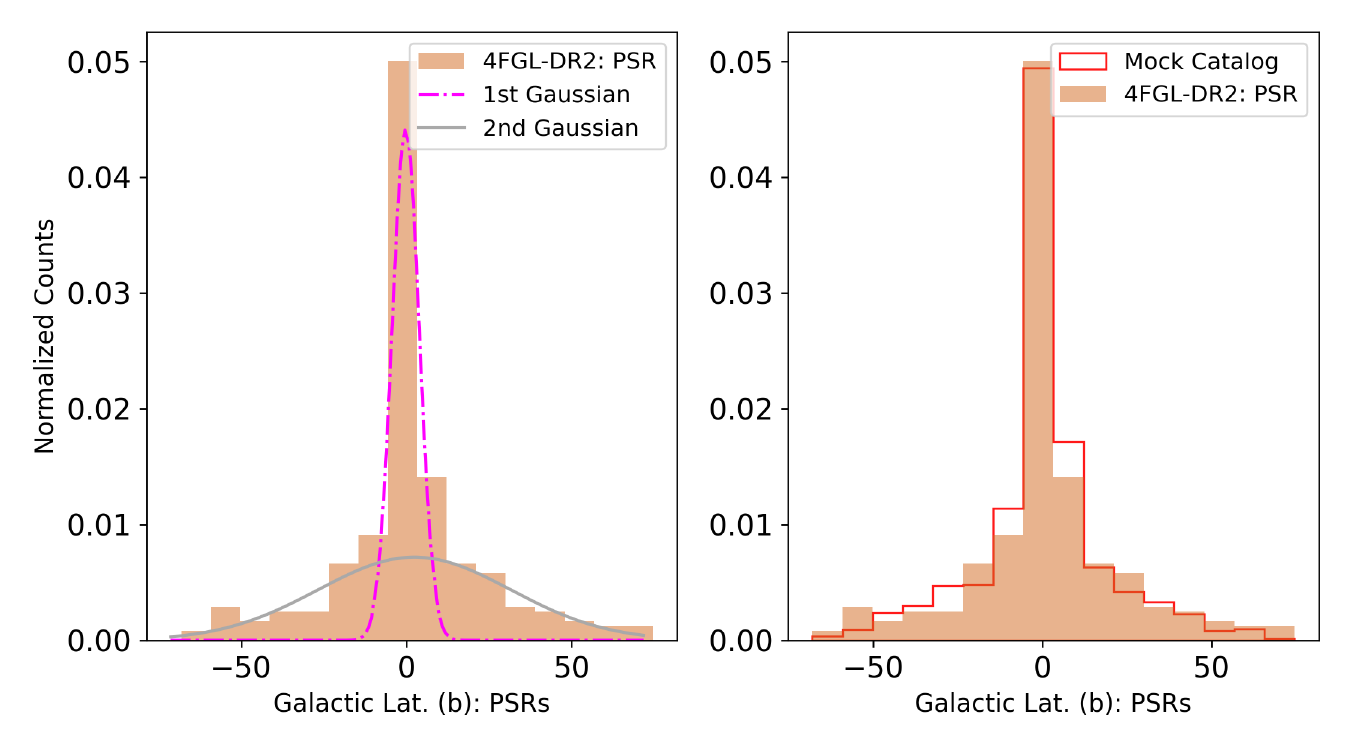}
    \includegraphics[width=0.96\columnwidth]{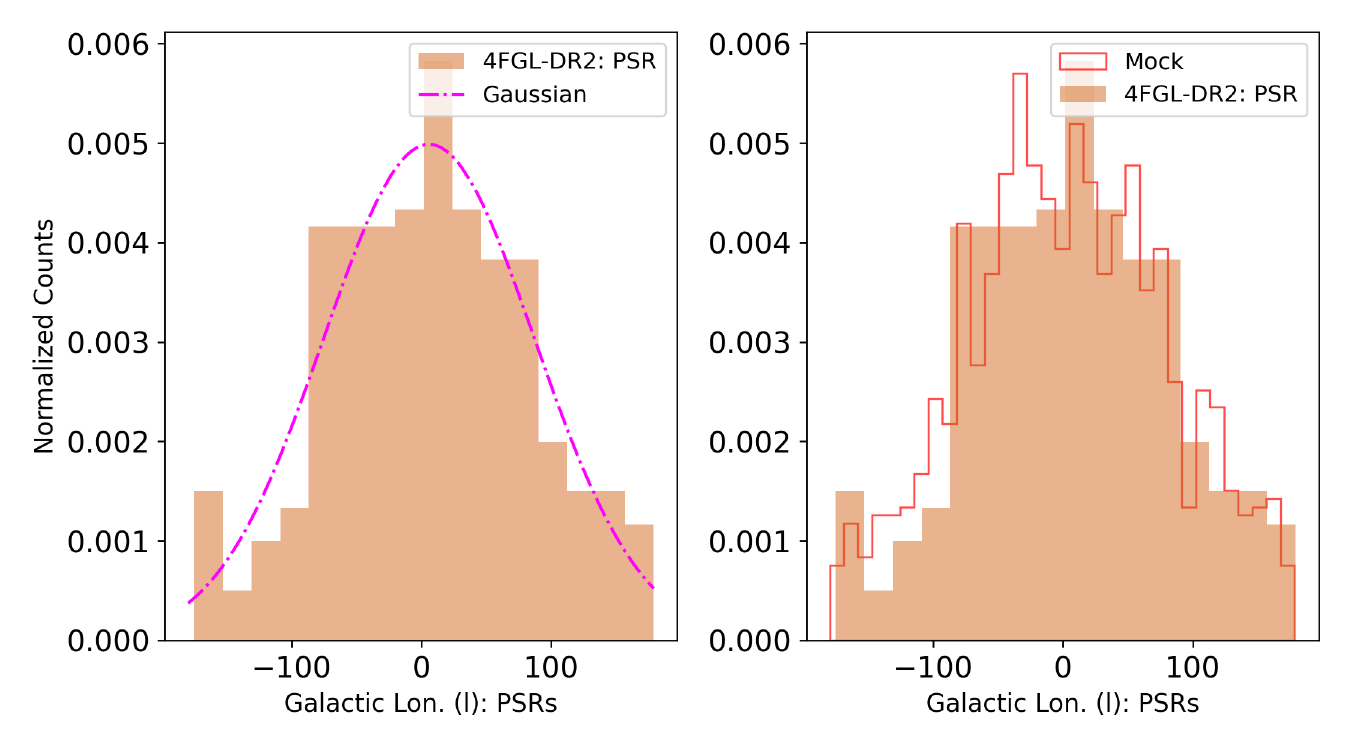}
    \caption{\small{Top panels: Galactic latitude distribution of PSR sources in the 4FGL-DR2 catalog modeled with two Gaussian components. The left panel shows the fitted model, while the right panel shows a realization sampled from the resulting distribution. Bottom panels: Galactic longitude distribution of PSR sources modeled with a single Gaussian component (left) together with a sampled realization from the distribution (right).}
    \label{fig:PSR-Dist-Spatial}}    
\end{figure}

The PSR spatial distribution was modeled separately in latitude and longitude. The latitude distribution was described as the sum of two Gaussian components centered on the Galactic plane, representing the young and millisecond pulsar populations. The two components, together with their sum, are shown in the upper panels of Figure~\ref{fig:PSR-Dist-Spatial}. The longitude distribution was modeled with a single Gaussian function, shown in the lower panels of Figure~\ref{fig:PSR-Dist-Spatial}. In both cases, the distributions obtained for the simulated catalog are compared with those of the 4FGL-DR2 PSR population.

\section{Detection and Localization}
\label{app:det-and-loc}

\begin{figure*}
    \centering
    \includegraphics[width=0.5\linewidth]{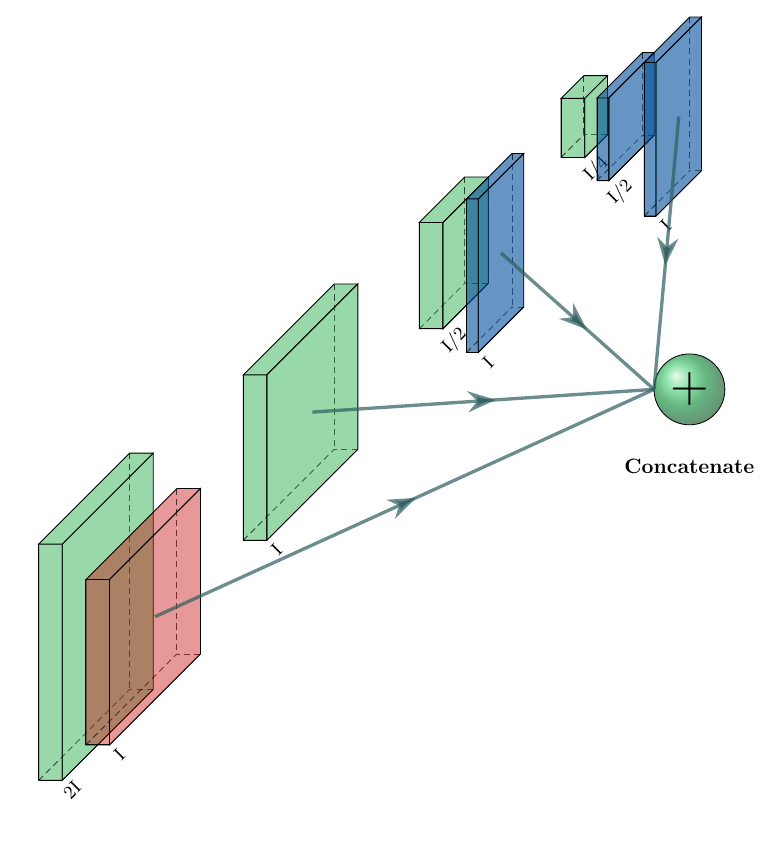}
    \caption{\small{Preprocessing layer introduced to extend the original U-Net architecture into the MIUN framework (Figure~\ref{fig:MIUN-arch}). Green boxes indicate convolution layers, blue boxes represent upsampling layers, and the red box denotes a convolution layer with stride used for downsampling.}}
    \label{fig:pre-processing}
\end{figure*}

\begin{figure*}
    \centering
    \includegraphics[width=0.8\linewidth]{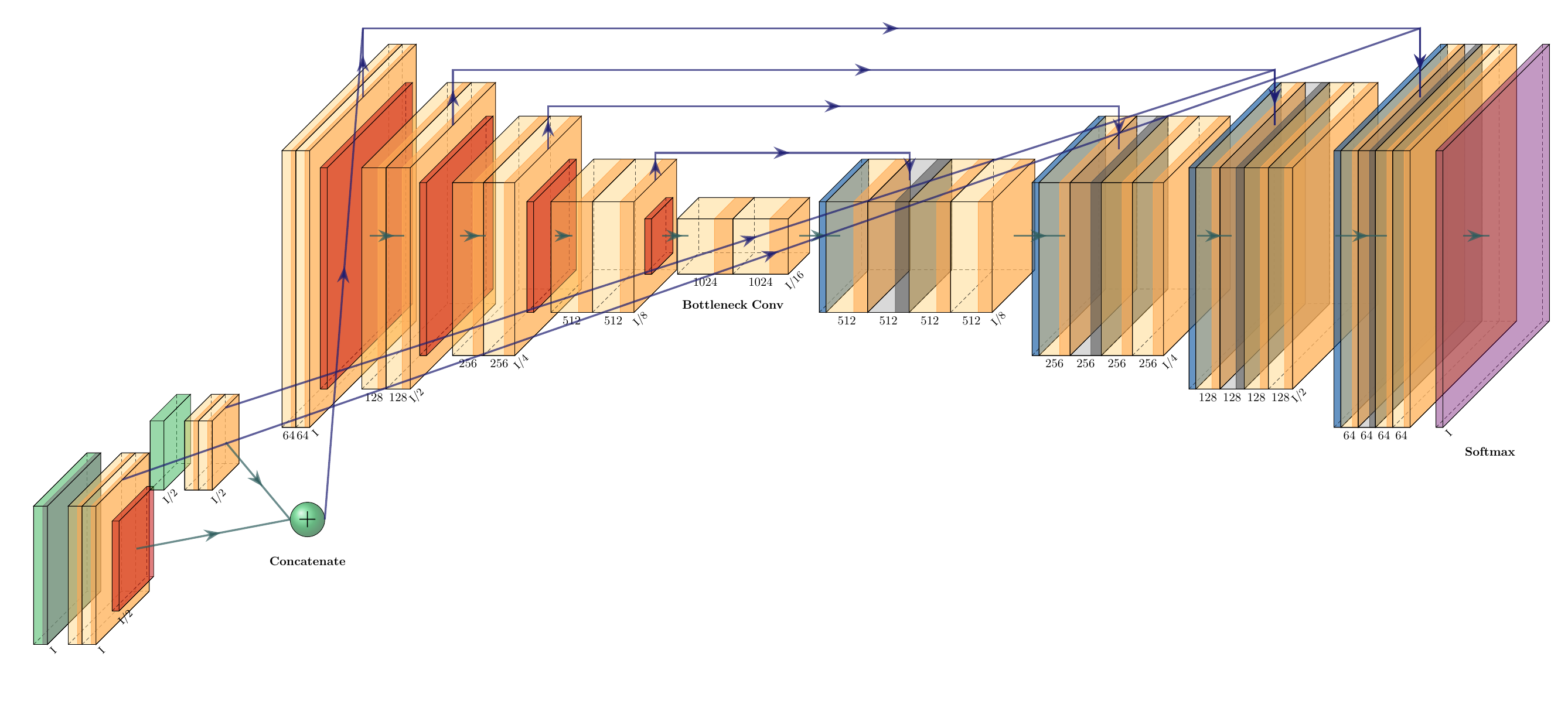}
    \caption{\small{Schematic representation of the full MIUN architecture. The only modification relative to the original U-Net is the preprocessing layer shown in Figure~\ref{fig:pre-processing}.}}
    \label{fig:MIUN-arch}
\end{figure*}

In this appendix, we provide additional details on the extension of the U-Net architecture developed for the detection of $\gamma$-ray sources, as well as on the source-localization procedure based on the Laplacian of Gaussian (LoG) method. A schematic representation of the architecture is shown in Figure~\ref{fig:MIUN-arch}. The main modification relative to the original U-Net is the preprocessing layer illustrated in Figure~\ref{fig:pre-processing}.
 
The network takes as input images from different energy bins with four spatial resolutions of the form $(I', I', 1)$, where $I' = 32, 64, 128,$ or $256$ pixels (Table~\ref{tab:IRFs}). These images are first processed by the preprocessing block, whose output has a common shape $(I, I, 6)$ with $I=128$. 

As shown in Figure~\ref{fig:pre-processing}, input images with size $2I$ are downsampled to $I$ using TensorFlow \textrm{Conv2D} layers with stride 2. These layers apply learned convolutional kernels over local pixel neighborhoods while simultaneously reducing the spatial dimensions. Conversely, images with sizes $I/2$ and $I/4$ are upsampled to size $I$ using the TensorFlow \textrm{UpSampling2D} layer, which performs nearest-neighbor interpolation. Images already matching size $I$ remain unchanged. The resulting feature maps from all energy bins are then concatenated to form an input tensor of shape $(I, I, N)$ for the conventional U-Net, where $N$ is the number of energy bins. The final output of the network is a segmentation map of shape $(128,128,1)$. Here, we also highlight that while in the original U-Net, the convolution layers were unpadded, leading to progressive spatial reduction at each layer, we use the same padding throughout all convolutional blocks. This helps to preserve the spatial dimensions across convolutional layers within each resolution stage, simplifying skip connection alignment.

This architecture allows spectral and spatial information from all energy bins to be incorporated simultaneously. The network therefore learns a direct mapping from multi-energy photon-count images to a single source-segmentation map with an angular resolution of $\sim0.1^{\circ}$ (Table~\ref{tab:IRFs}).

The MIUN network was trained using the Adam optimizer with a starting learning rate of $1\times 10^{-3}$, a batch size of 32 and maximum epochs set to 250. Model performance during training was tracked via validation loss and the Dice Coefficient. To prevent overfitting and ensure convergence, we applied a dynamic learning rate schedule that reduced the learning rate by a factor of 0.8 following 5 consecutive epochs without improvement in validation loss. Early stopping with a patience of 8 epochs was implemented, and the best model weights were stored that produced minimum validation loss.

After obtaining the segmentation maps, we localize sources using the LoG blob-detection algorithm implemented in the \texttt{blob\_log} function of the \textrm{scikit-image} package \citep{scikit-image}\footnote{https://scikit-image.org/}. Two important hyperparameters of the LoG algorithm are the Gaussian width $\sigma$ and the overlap threshold. The parameter $\sigma$ controls the characteristic size of detectable blobs, with smaller (larger) values favoring compact (extended) structures. The overlap parameter suppresses duplicate detections by removing one blob when the overlap area between two blobs exceeds a specified fraction.

\begin{figure*}
    \centering
    \includegraphics[width=0.9\columnwidth]{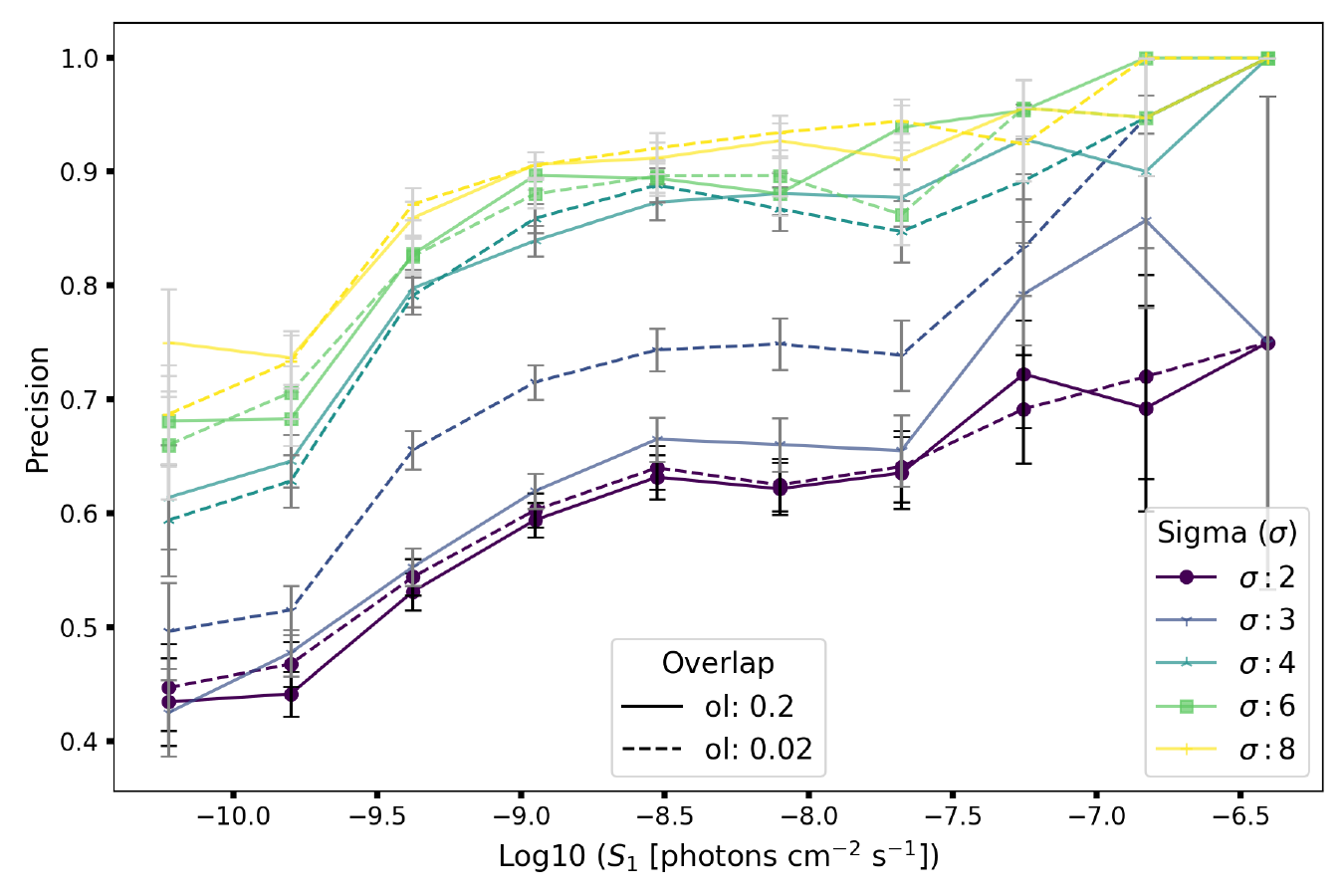}
    \includegraphics[width=0.9\columnwidth]{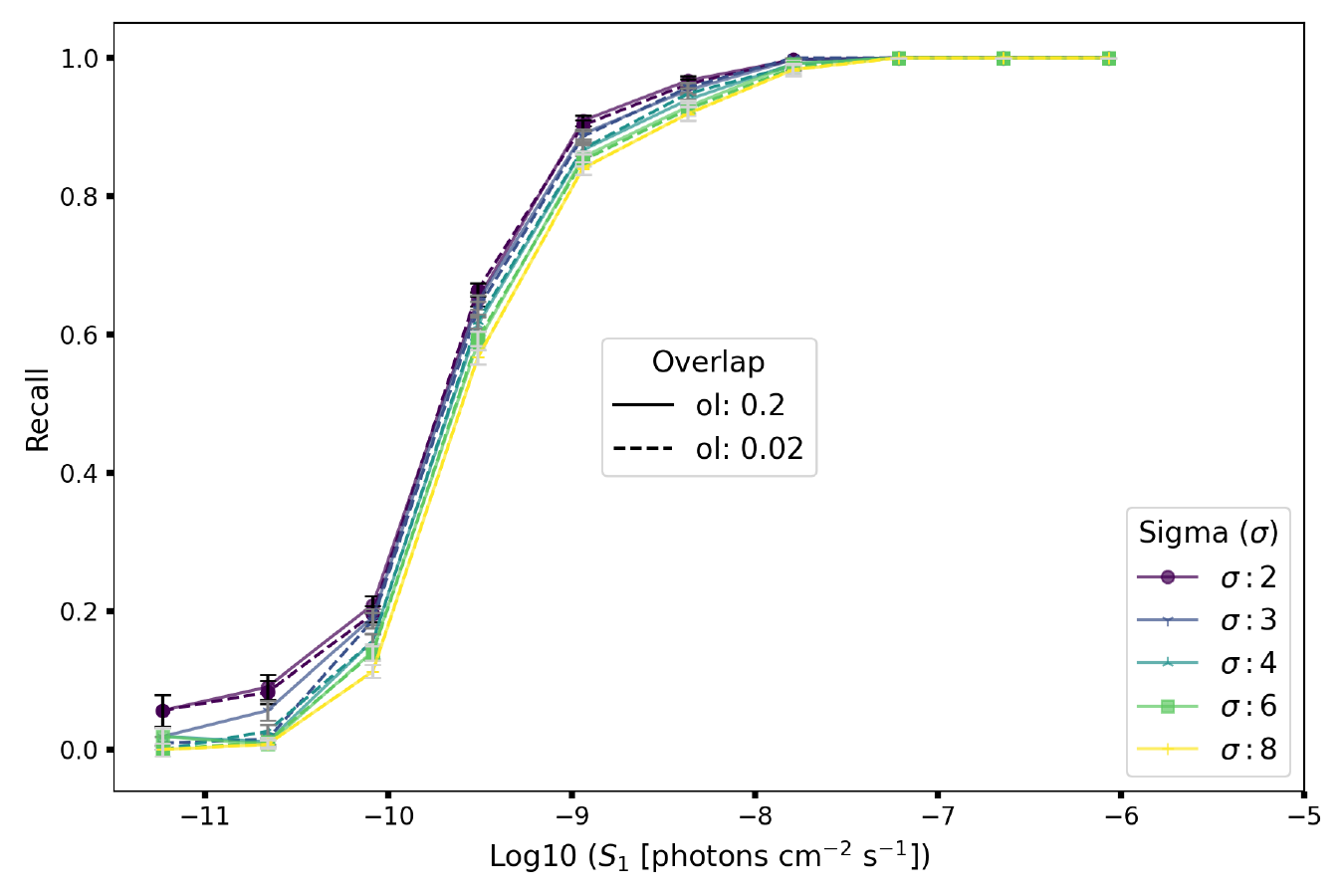}
    \caption{\small{Precision and recall as functions of the predicted flux $S_1$ for different choices of the LoG parameters $\sigma$ and overlap (\textrm{ol}).}
    }
    \label{fig:sigma-overlap-pre-rec}
\end{figure*}

To optimize these parameters, we evaluated the precision and recall as functions of $S_1$ for simulated datasets using different combinations of $\sigma$ and overlap values (Figure~\ref{fig:sigma-overlap-pre-rec}). While the recall remains relatively stable, the precision depends strongly on the parameter choice. In particular, very small values of $\sigma$ tend to produce multiple false-positive detections around true sources, thereby reducing the precision. We adopt $\sigma=6$ pixels and an overlap threshold of $0.2$ throughout this work. The chosen value $\sigma=6$ corresponds to an angular scale of approximately $0.6^{\circ}$, comparable to the 68\% containment radius of \textrm{FRONT}-converting LAT events at $\sim1$~GeV.

\subsection{Source Confusion: Nearby Sources in Detection}
To evaluate the robustness of the MIUN + Laplacian-of-Gaussian (LoG) framework in crowded regions, we performed a dedicated Monte Carlo (MC) study of source confusion and close-pair separation.
Additional synthetic sources were injected into simulated patches near existing astrophysical sources. To test the sensitivity of the detection and localization pipeline to relative source brightness, the injected MC sources were generated under two conditions:
(i) with the same total counts as the nearby source, and
(ii) with fluxes systematically varied with respect to the nearby source by a factor of 10.
These configurations probe cases of neighboring injected sources of varying brightness. 

\begin{figure}
    \centering
    \includegraphics[width=0.93\linewidth]{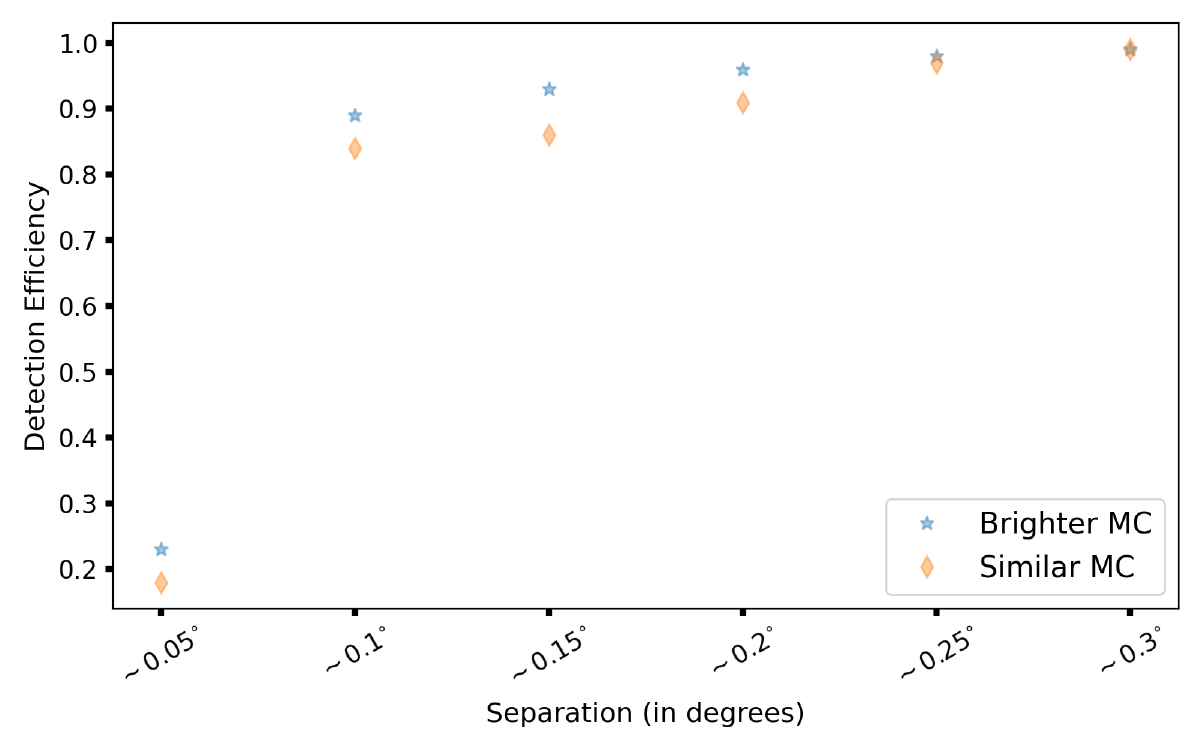}
    \caption{\small{Detection efficiency of the combined MIUN + LoG pipeline as a function of source separation for different relative brightnesses of the injected MC source and the nearby astrophysical source. A successful detection requires both sources to be recovered.}}
    \label{fig:source-sep-MC}
\end{figure}

The overall performance of the detection and localization pipeline for this nearby-source injection scenario is summarized in Figure~\ref{fig:source-sep-MC}. This figure shows the detection efficiency as a function of angular separation for the different brightness configurations. A detection is counted only when both the injected MC source and the nearby astrophysical source are successfully recovered.

\section{Classification: True vs FPs}\label{sec:app-classification}

\begin{figure*}
    \centering
    \includegraphics[width=0.9\columnwidth]{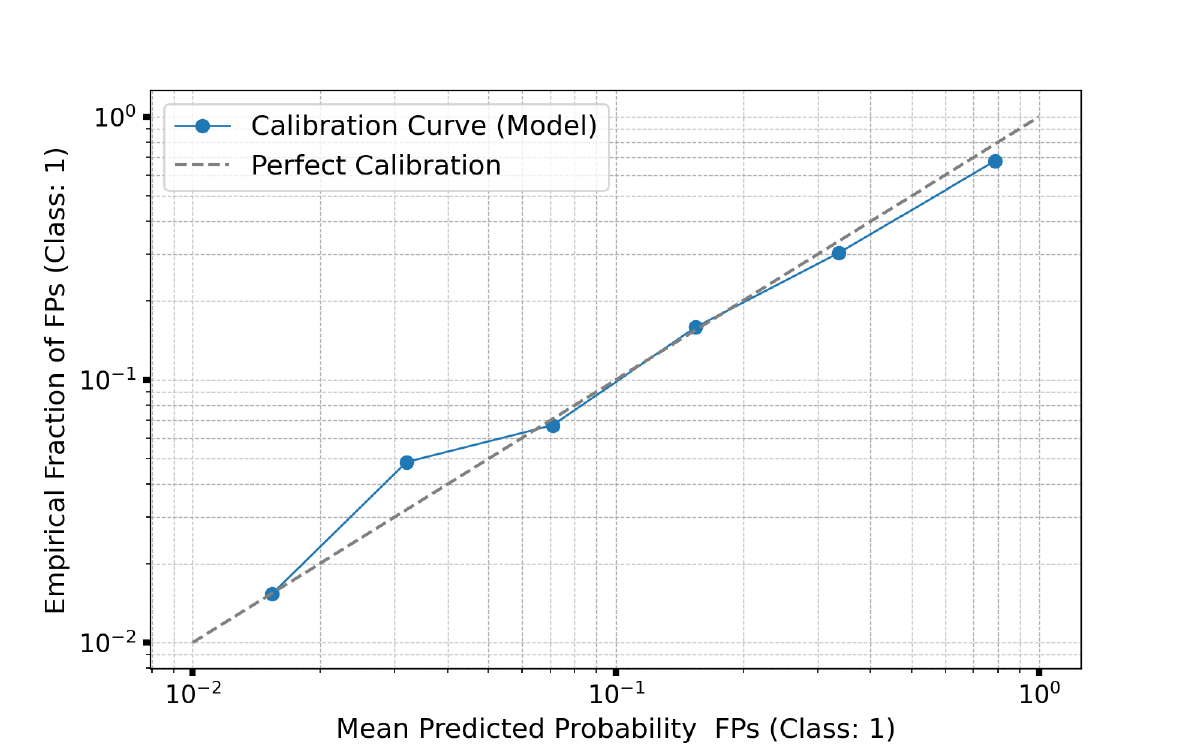}
    \includegraphics[width=0.9\columnwidth]{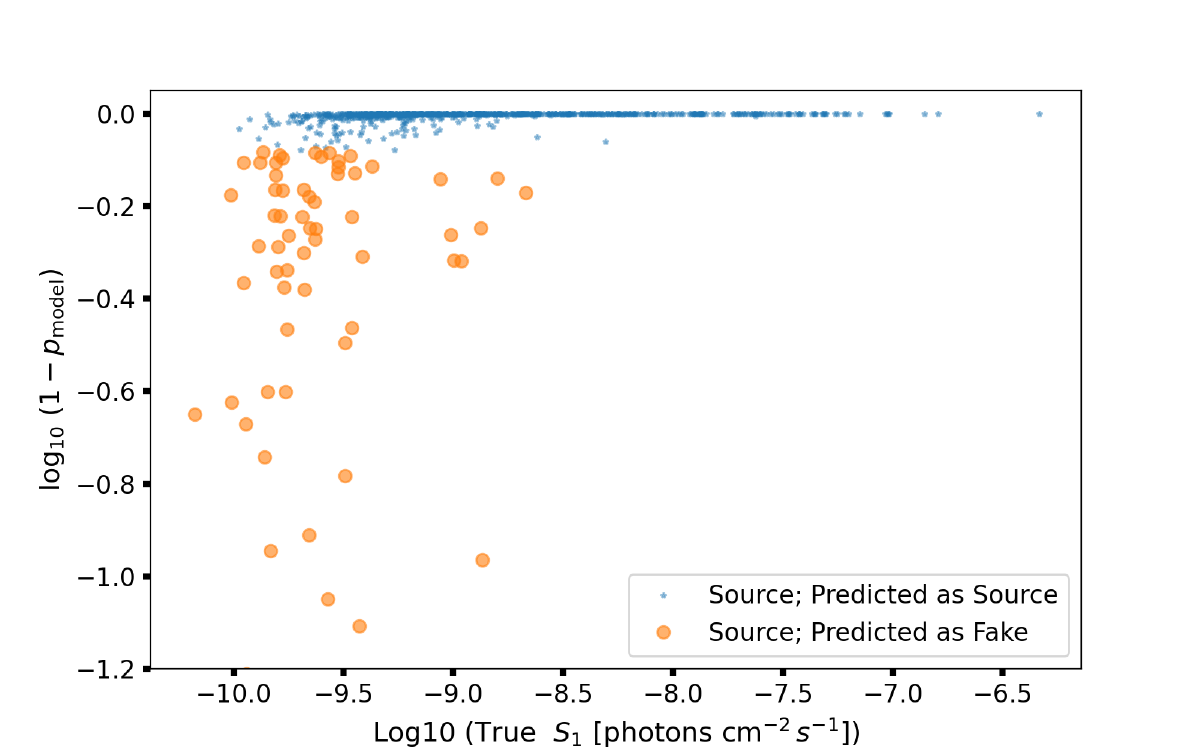}
    \caption{\small{Left: Calibration curve for FP-source probabilities over the full sky. The blue points show the empirical FP fraction $\left(\frac{FP}{FP+TP}\right)$ in each predicted-probability bin, while the diagonal dotted line indicates perfect calibration. Right: $\log(1-p_{\mathrm{model}})$ as a function of the true source flux $S_1$, where $p_{\mathrm{model}}$ is the network-predicted FP-source probability. Correctly classified TP sources cluster near zero (blue points), while misclassified TP sources are shown in orange.}   
    }
    \label{fig:class-calib-sigmoid}
\end{figure*}

In this appendix, we provide additional details on the classification of TP and FP sources within the source characterization module.
Because the training dataset is imbalanced, containing only $\sim 19\%$ FP sources, the optimal classification threshold was determined by maximizing the TPR and minimizing FPR. Since the primary goal of this stage is to suppress false positives, we evaluate the reliability of the predicted FP-source probabilities through a calibration analysis.
To do this, we bin the sources in log(predicted FP probability) and compute the empirical fraction of FPs in each bin.
The left panel of Figure~\ref{fig:class-calib-sigmoid} shows the calibration curve for the predicted FP-source probability. 
The model predictions are in reasonably good agreement with the ideal calibration relation (diagonal line), indicating that the predicted probabilities are well calibrated and statistically meaningful. For instance, as illustrated by the first point in the plot, a predicted FP probability of approximately $0.015$ corresponds to an observed FP fraction of about $1.5\%$. At predicted probabilities above $0.1$, the relative difference between the predicted and observed FP fractions is reduced to approximately $10$--$20\%$.
In the right panel of Figure~\ref{fig:class-calib-sigmoid}, we show the raw sigmoid scores ($p_{\text{model}}$ in the figure) as a function of the true source flux $S_1$. The network tends to misclassify faint TP sources as FP, while bright TP sources are generally assigned high confidence scores. This trend emerges naturally, even though no explicit calibration procedure or flux-related information was provided during training.

\section{Background Model Independence}

\begin{figure}
    \centering
        \includegraphics[width=0.93\linewidth]{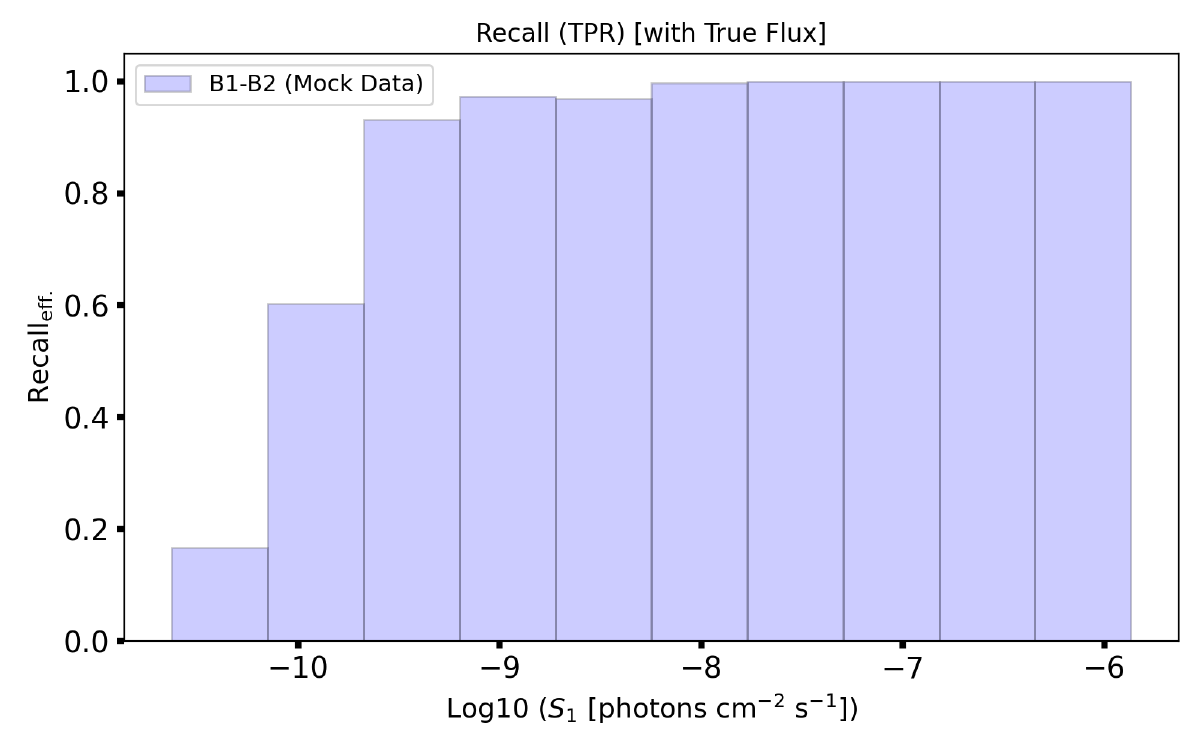}    
    \caption{Same as Figure~\ref{fig:bkg-sys-Pre-Rec-Eff}, but shown as a function of the true simulated flux $S_1$.}
    \label{fig:Bkg-IEM-Recall-Tr-Flx}
\end{figure}

In Section~\ref{sec:bkg-syst}, we introduced the effective precision and recall, $\text{Precision}_{\text{eff}}$ and $\text{Recall}_{\text{eff}}$, as functions of the reconstructed flux $S_1$ using the flux-prediction network (Figure~\ref{fig:bkg-sys-Pre-Rec-Eff}). In this appendix, we instead show $\text{Recall}_{\text{eff}}$ as a function of the true simulated flux $S_1$ in Figure~\ref{fig:Bkg-IEM-Recall-Tr-Flx}.
The effective recall decreases to $\sim 60\%$ near the detection threshold,
$S_1 \approx 2\times10^{-10}\,\mathrm{ph\,cm^{-2}\,s^{-1}}$,
which is expected in the lowest-flux regime.
This behavior is partly driven by flux migration effects: bright sources may have their flux underestimated and shift into lower-flux bins, while faint sources can occasionally scatter upward. 

Since the true $S_1$ values are unavailable for real data, our analysis relies on the flux-prediction network. Improved flux calibration and future refinements of the network architecture may help reduce these migration effects.

\section{1FDL Source List Creation}

\subsection{1FDL Source List Parameters}
The 1FDL source list contains both the quantities derived in this work and selected columns from the 4FGL-DR2 catalog. Table~\ref{tab:1fdl-cols} summarizes the columns specifically produced during the 1FDL list creation.
For each source, we provide the Galactic coordinates (1FDL\_LON, 1FDL\_LAT) obtained from the detection and localization pipeline (Section~\ref{sec:detection}) and refined with the deep-ensemble method (Section~\ref{sec:CharModule}). The localization uncertainties are used to derive positional error ellipses.
We also provide the predicted photon flux and $1\sigma$ uncertainty between 1 and 100 GeV (1FDL\_Flux1000, 1FDL\_Unc\_Flux1000), analogous to the \texttt{Flux1000} and \texttt{Unc\_Flux1000} quantities in the 4FGL catalog. In addition, 1FDL list contains the calibrated class probability (1FDL\_Class\_Prob) for all the candidates that are already classified as True Positives.
Finally, for sources associated with 4FGL-DR2 entries, we include relevant catalog information such as the 4FGL source name, source class, and corresponding multiwavelength association for convenient cross-identification.

\begin{table*}
    \caption{Description of the entries provided in the 1FDL source list. The online version additionally includes selected columns from the 4FGL-DR2 catalog.}
    \centering
    \begin{tabular}{cccc}
         \hline \hline
         Column & Format & Unit & Description \\
         \hline
         \\
         1FDL\_Name & 18A& --- & Source name 1FDL JHHMM.m+DDMM \\
         \\
         1FDL\_LON & E& deg. & Galactic Longitude \\
         \\
         1FDL\_LAT & E & deg. & Galactic Latitude \\
         \\
         1FDL\_RA & E & deg. & Right Ascension  \\
         \\
         1FDL\_DEC & E & deg. & Declination \\
         \\
         1FDL\_Conf\_Radius & E & deg. & Location Error Radius $(68\%)$ \\
         \\
         1FDL\_Flux1000 & E & $\text{cm}^{-2}\, \text{s}^{-1}$ & Photon flux from 1~GeV to 100~GeV \\
         \\
         1FDL\_Unc\_Flux1000 & E & $\text{cm}^{-2}\, \text{s}^{-1}$ & $1\sigma$ error on 1FDL\_Flux1000 \\
         \\
         1FDL\_Class\_Prob & D & --- & Calibrated Source Class Probability\\
         \\
         DR2\_Source\_Name & 18A & --- & Source Name of Original DR2 Source\\
         \\
         DR2\_CLASS1 & 5A & --- & Class designator from original DR2 Catalog\\
         \\
         DR2\_ASSOC1 & 28A & --- & Name of identified or associated source from DR2 Catalog\\
         \\
         \hline
    \end{tabular}
    
    \label{tab:1fdl-cols}
\end{table*}

\subsection{Flux Prediction on Real Data}\label{app:flux-prediction-DR2}

Here we present a detailed performance evaluation of the flux reconstruction network on real data. As we have shown a comparison of the recall rate on real data as a function of DR2 $\log S_1$ bins in Figure \ref{fig:Recall-DR2-ASID-FL8Y-DR4} to demonstrate the model robustness against background systematics, here, we highlight the flux prediction network performance on the real data. 

\begin{figure}
    \centering
    \includegraphics[width=0.94\linewidth]{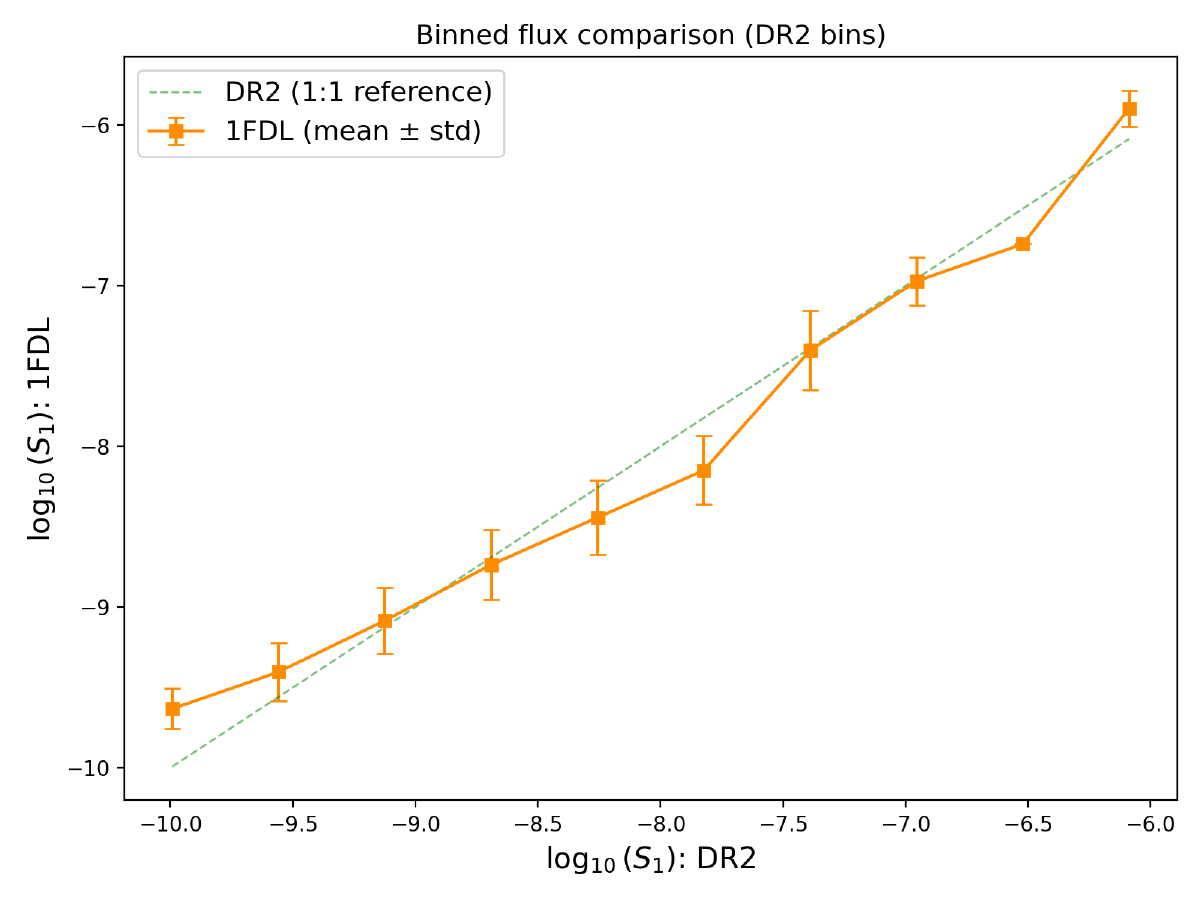}
    \caption{\small{Binned $S_1$ (in DR2 $\log (S_1)$ bins) for the associated sources are compared between DR2 $S_1$ and 1FDL predicted $S_1$. For every $\log (S_1)$ bin, the 1FDL population mean and $1\sigma$ standard deviation are shown.}}
    \label{fig:binned-flux-DR2-FDL}
\end{figure}

First, we show in Figure \ref{fig:binned-flux-DR2-FDL} a binned flux comparison from our method against the standard 4FGL-DR2 flux bins. For all the 1FDL sources associated with 4FGL-DR2, we calculate the population mean and $1\sigma$ standard deviation of the 1FDL sources for each of these flux bins. Figure \ref{fig:binned-flux-DR2-FDL} highlights that the reconstructed flux from the flux reconstruction network tracks the diagonal line (DR2 flux), and the small deviations are mostly within the error bars.

\begin{figure}
    \centering
    \includegraphics[width=0.94\linewidth]{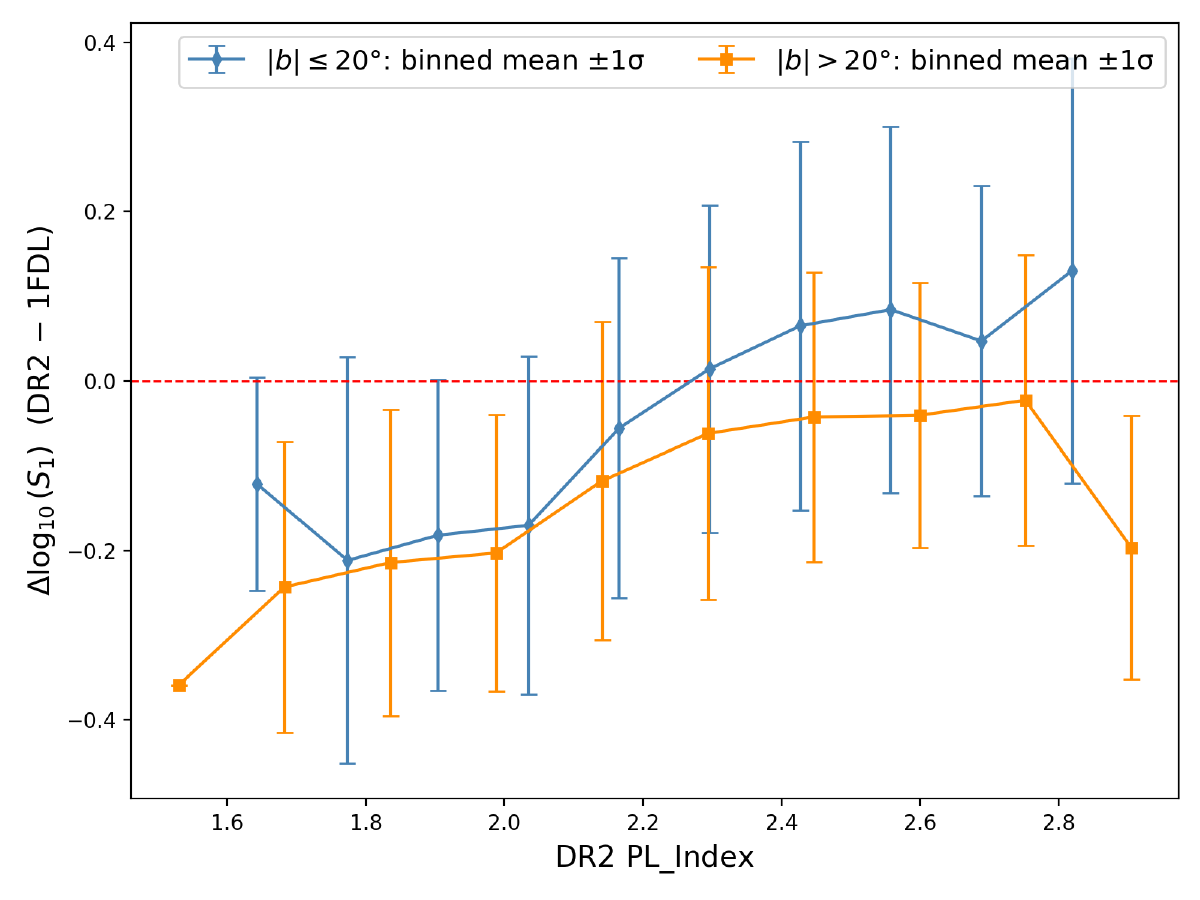}
    \caption{\small{Log $S_1$ residual between DR2 and 1FDL is plotted against the power-law index (\textrm{`PL\_Index'}) of the DR2 sources. The errorbars represent the binned standard deviation of the residual within each \textrm{`PL\_Index'} bin and are shown separately for sources away (close) from the Galactic plane in orange(blue). }}
    \label{fig:flux-resid-vs-PLIndex}
\end{figure}

In Figure \ref{fig:flux-resid-vs-PLIndex}, we show the flux residual between DR2 $S_1$ and 1FDL $S_1$, $(\Delta \log _{10} (S_1) ~(\text{DR2}-\text{1FDL})$, as a function of the DR2 Power-Law index (\textrm{PL}\_\textrm{Index}), for all the associated sources. As shown, the residual is relatively small for DR2 sources with \textrm{PL}\_\textrm{Index} between 2.1 and 2.4, whether they lie close to the plane $(|b|<20^{\circ})$, or away from the galactic plane, compared to sources with very hard spectra. Compared to the likelihood method, where first a power-law index is obtained by fitting the source spectrum and then calculating the source flux, here our inputs to the DNN are localized images at various energy bins. Thus, it becomes difficult for the model to predict fluxes of sources with softer (harder) spectra, which contribute negligibly at higher (lower) energy bins.   

\begin{figure}
    \centering
    \includegraphics[width=0.94\linewidth]{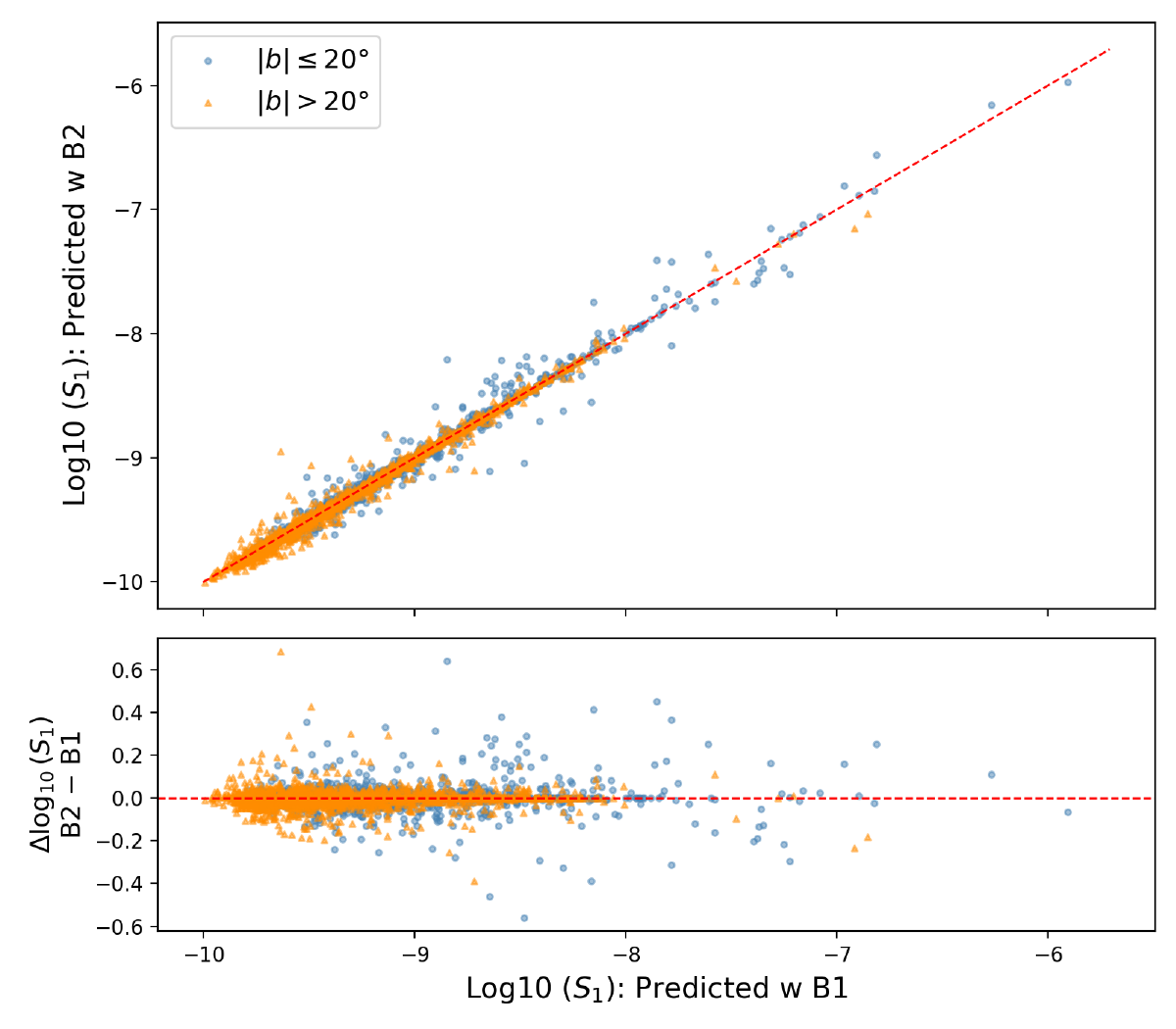}
    \caption{\small{Predicted $S_1$ on the real data obtained from a model trained on B1 IEM is compared against the predictions from a model trained on B2 IEM. The residual between these predictions is shown in the panel below. Galactic (extragalactic) sources are marked with orange and blue colors.}}
    \label{fig:b1-b2-real-data-flux}
\end{figure}

Finally, we also show the consistency in flux prediction on real data, irrespective of the background model used for training the model in Figure \ref{fig:b1-b2-real-data-flux}. The top panel of Figure \ref{fig:b1-b2-real-data-flux} shows the scatter plot of $\log _{10}(S_1)$ predicted using model B2 against the predictions using model B1. They tightly follow the diagonal line, both for sources away from the plane $|b|>20^{\circ}$, (shown with orange points), and close to the galactic plane (shown in blue). The corresponding residual plot (bottom panel), defined as $\Delta \log_{10} (S1) (B2-B1)$, shows that residuals are indeed close to zero. Relatively higher scatter for sources with low fluxes and specifically for sources affected by the IEM (near the Galactic plane ($|b|<20^{\circ}$)) is also evident in the residual plot. 

\subsection{Bright Unassociated DR2 Sources}

\begin{figure*}
    \centering
    \includegraphics[width=0.96\linewidth]{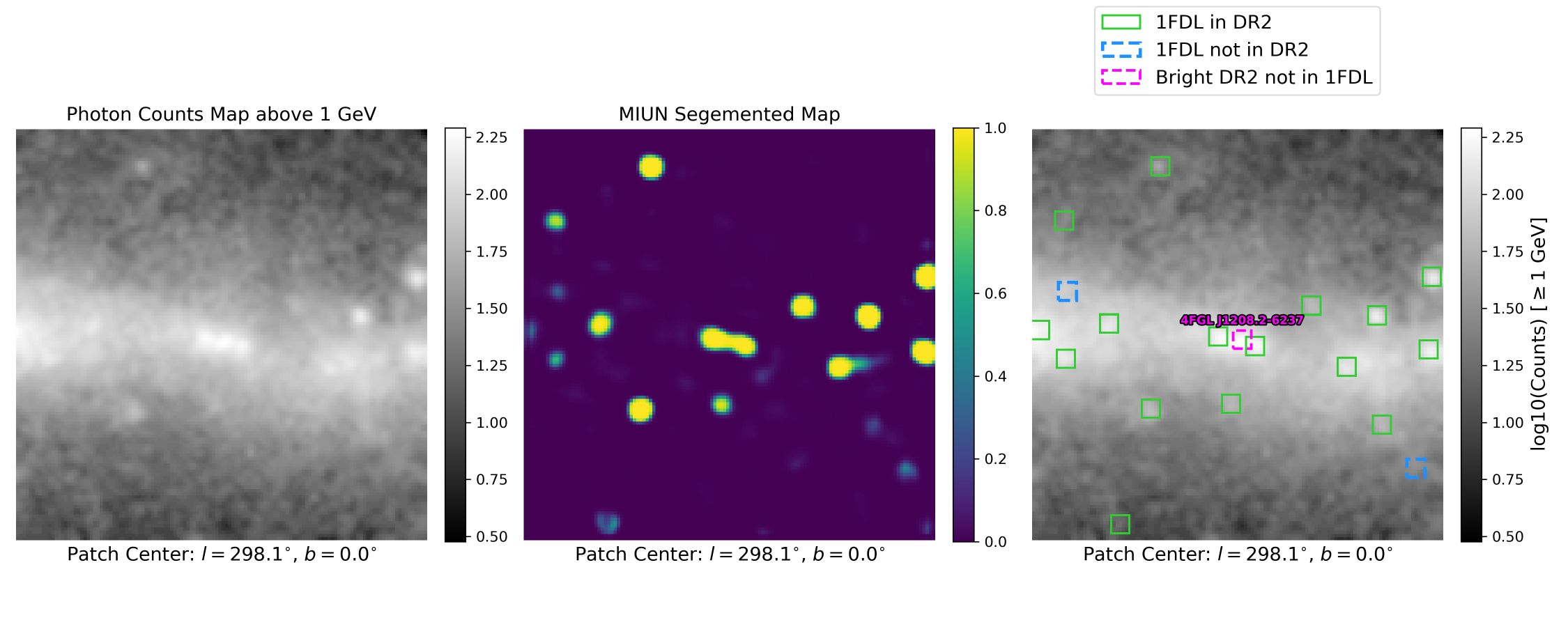}
    \caption{\small 
    Bright DR2 sources (\textrm{Signif\_Avg}>20) without counterparts in the 1FDL list are shown in magenta. The patch sizes are $\sim 10^{\circ} \times 10^{\circ}$. 
    A bright DR2 source is between two 1FDL sources.
    The segmentation map (middle panel) has difficulty in producing distinct blobs in a source-dense region with bright IEM (left panel).}
    \label{fig:unassoc-bright-2patch}
\end{figure*}

Most bright (\textrm{`Signif\_Avg'}$\geq$20) unassociated DR2 sources are located in crowded regions with strong diffuse $\gamma$-ray emission (Figure \ref{fig:DR2-Unassoc-HSignif}).  
In these strong diffuse emission dominated environments, the MIUN segmentation step may fail to produce well-separated source masks (shown in the segmented MIUN map in Figure \ref{fig:unassoc-bright-2patch}), which in turn affects the subsequent LoG centroid reconstruction. The highlighted bright (\textrm{`Signif\_Avg'}=28.049) DR2 pulsar, 4FGL J1208.2-6237, in Figure \ref{fig:unassoc-bright-2patch}, lies between two other bright DR2 sources: 4FGL J1203.9-6242 (\textrm{`Signif\_Avg'}=34.506) and 4FGL J1213.3-6240e (\textrm{`Signif\_Avg'}=34.594); both of which are associated to the 1FDL list. 

\end{appendix}

\bibliographystyle{aa}
\bibliography{aa-ref}


\end{document}